\documentclass[fleqn,usenatbib]{mnras}
\usepackage{hyperref}  

\usepackage{etoolbox}
\makeatletter
\newcount\c@additionalboxlevel
\newcount\c@maxboxlevel
\patchcmd\@combinedblfloats{\box\@outputbox}{%
  \stepcounter{additionalboxlevel}%
  \box\@outputbox
}{}{\errmessage{\noexpand\@combinedblfloats could not be patched}}

\AtBeginShipout{%
  \ifnum\value{additionalboxlevel}>\value{maxboxlevel}%
    \typeout{Warning: maxboxlevel might be too small, increase to %
      \the\value{additionalboxlevel}%
    }%
  \fi 
  \@whilenum\value{additionalboxlevel}<\value{maxboxlevel}\do{%
    \typeout{* Additional boxing of page `\thepage'}%
    \setbox\AtBeginShipoutBox=\hbox{\copy\AtBeginShipoutBox}%
    \stepcounter{additionalboxlevel}%
  }%
 \setcounter{additionalboxlevel}{0}%
}
\makeatother

\usepackage{amsmath}
\usepackage{amssymb}
\usepackage{cite}
\usepackage{natbib}
\usepackage{graphicx}
\usepackage{xcolor}
\DeclareMathOperator{\arctanh}{arctanh}
\usepackage[T1]{fontenc}
\usepackage{ae,aecompl}

\def\simlt{\lower.5ex\hbox{$\; \buildrel < \over \sim \;$}}
\def\simgt{\lower.5ex\hbox{$\; \buildrel > \over \sim \;$}}
\def\msun{M_\odot}

\let\urshowkeys=\showkeys \def\showkeys{\needspace{5ex}\urshowkeys}

\title[Zooming into the Cosmic Horseshoe]{%
Zooming into the Cosmic Horseshoe: \\new insights on the lens profile and the source shape
}

\author[Bellagamba, Tessore, Metcalf]{%
Fabio Bellagamba,$^1$ 
Nicolas Tessore$^{1,2}$ and
R. Benton Metcalf$^1$ \\
$^1$Department of Physics and Astronomy, Universit\`a di Bologna, viale Berti Pichat 6/2, 40127 Bologna, Italy \\
$^2$Jodrell Bank Centre for Astrophysics, University of Manchester, Alan Turing Building, Manchester M13 9PL, UK}

\pubyear{2016}

\begin{document}
\label{firstpage}
\pagerange{\pageref{firstpage}--\pageref{lastpage}}
\maketitle

\begin{abstract}
The gravitational lens SDSS J1148+1930, also known as the Cosmic Horseshoe, is one of the biggest and of the most detailed Einstein rings ever observed.
We use the forward reconstruction method implemented in the lens fitting code \textsc{Lensed} to investigate with great detail the properties of the lens and of the background source.
We model the lens with different mass distributions, focusing in particular on the determination of the slope of the dark matter component.
The inherent degeneracy between the lens slope and the source size can be broken when we can isolate separate components of each lensed image, as in this case.
For an elliptical power law model, $\kappa(r) \sim r^{-t}$, the results favour a flatter-than-isothermal slope with a maximum-likelihood value $t = 0.08$.
Instead, when we consider the contribution of the baryonic matter separately, the maximum-likelihood value of the slope of the dark matter component is $t = 0.31$ or $t = 0.44$, depending on the assumed Initial Mass Function.
We discuss the origin of this result by analysing in detail how the images and the sources change when the slope $t$ changes.
We also demonstrate that these slope values at the Einstein radius are not inconsistent with recent forecast from the theory of structure formation in the $\Lambda$CDM model.
\end{abstract}

\begin{keywords}
gravitational lensing: strong -- galaxies: individual: Cosmic Horseshoe -- galaxies: structure -- dark matter 
\end{keywords}

\section{Introduction}

The slope of the mass distribution within galaxy halos has been a long-standing prediction of the $\Lambda$CDM model since \citet*[NFW]{1997ApJ...490..493N} found that the halos in collisionless dark matter $N$-body simulations follow a universal average profile. There have been small modifications \citep{2004MNRAS.349.1039N}, but it remains a robust prediction of the model today.
Measuring this slope from observations has proven very difficult owing to the need for some clear tracer of the density and accurate predictions for the behaviour of baryons within galaxies \citep{2014Natur.506..171P}.

At large radii, the NFW prediction is consistent with weak galaxy-galaxy lensing studies \citep{2008JCAP...08..006M}.  Here the tangential shearing of background galaxies is correlated with their proximity to foreground galaxies.  Statistically the large radius behavior  $\rho \propto r^{-3}$ is found.  At small radii there are fewer chance alignments and the lensing begins to cause nonlinear distortions to the images.  Here averaging the distortion from many lenses fails to provide meaningful statistical constraints on the mass distribution. Instead lens models need to be fit to each lenses separately.

Strong gravitational lens modeling provides a unique way of probing the inner regions of dark matter halos, but it is encumbered by a near degeneracy that makes it very difficult to measure the slope of the dark matter profile using lensing data alone.
In most cases, an Einstein ring (a galaxy image distorted into a partial or total ring around the center of a lens galaxy) provides a very good measure of the total mass within the ring, but does not constrain the slope of the mass profile at the ring.
Recent studies have combined dynamical measures of the mass within the visible lens galaxy with lensing data \citep{2010ApJ...724..511A, Oguri11042014}.
In this way a single power-law density model can be fit to the total mass, dark matter and baryons.
These studies have found that the total mass is well fit by an isothermal profile ($\rho \propto r^{-2}$) out to several effective radii.
Stellar population synthesis modeling is typically too uncertain to put strong independent constraints on the mass in stars \citep{2016MNRAS.tmp..677L}.
In one unique case on galactic scales, SDSS J0946+1006 \citep{2008ApJ...677.1046G}, the existence of two Einstein rings makes it in principle possible to constrain a two component model to the lens \citep{2012ApJ...752..163S}.
Even in this case however, it is not clear if the contributions from stars and dark matter can be so easily separated.

The mass profile of nearby elliptical galaxies has also been constrained through pure dynamical modelling, with the conclusion that the total mass profile is close to isothermal within 4 effective radii, with some scatter in the slope \citep{2014MNRAS.437.3670C,2015ApJ...804L..21C}.
This is generally in agreement with the combined dynamics and lensing results.

Another approach has been to study the rotation curve of gas in dwarf galaxies that are believed to be dark matter dominated (e.g.\@ \citealp{1538-4357-447-1-L25,2007ApJ...663..948G}).
These studies show some evidence of a maximum density in the core of dark matter halos.
There have also been several studies that have constrained the slope of the mass profile using strong gravitational lensing and dynamics on galaxy cluster scales.
These studies have indicated that the dark matter profile has a core in clusters at radii ${\simlt}~30$~kpc \citep{2002ApJ...574L.129S,2013ApJ...765...25N}.
The existence and physical origin of such a core in the dark matter profile remains uncertain \citep{1538-4357-529-2-L69,0004-637X-561-1-35,Weinberg06102015,Laporte01082015}.

In this paper, we seek to constrain the slope of the dark matter profile at the Einstein ring radius where the projected surface density is highly dark matter dominated using detailed strong lensing modeling alone.
It has long been known that this is very difficult because changing the slope of the density is nearly degenerate with the total magnification of the source which cannot be determined independently.
However, as will be discussed below, it is possible to measure the slope using the relative radial magnifications of multiple resolved images, as pointed out by \citet{2012MNRAS.426..868S}.
\citet{2014A&A...564A.103S} have shown that degeneracies exist that make it nearly impossible to measure the radial slope in the general case.
However, if the space of lens models is restricted to one particular class, for example power-law profiles with elliptical symmetry, such a measurement is possible, as we will demonstrate.

A related point is that the magnification of the source is dependent on the slope of the mass profile at the Einstein radius and not the average slope within that radius.
As a result using a isothermal model when investigating the properties of the source might lead to significant errors.
We investigate this further in Section~\ref{sec:degeneracy}.

This paper is organised as follows.
The next Section describes the data and the object used in this study.  Section~\ref{sec:method} has a description of the method used to model the lens.
In Section~\ref{sec:results} we present the results of the modelling procedure for the source and lens model.
We discuss the implications of these results in Section~\ref{sec:discussion} in terms of the limitations of lens modeling, the CDM model and in comparison with previous results.
The paper is concluded in Section~\ref{sec:conclusion} with a summary of the main results.

When necessary, we use a $\Lambda \text{CDM}$ cosmology with $\Omega_\text{M}$~=~0.3, $\Omega_\Lambda$ = 0.7, and $h$ = 0.7.

\section{Data} \label{sec:data}
The Cosmic Horseshoe was discovered by \citet{2007ApJ...671L...9B} while searching the Sloan Digital Sky Survey (SDSS) for luminous red galaxies with multiple faint blue companions. 
The centre of the lens galaxy lies at ($11^h 48^m 33^s.15, 19^{\circ} 30^{\prime} 3^{\prime\prime}.5$).
The lens system features a nearly complete Einstein ring (${\sim}~300^\circ$) with a very large diameter (${\sim}~10^{\prime\prime}$), which corresponds to an enclosed mass of $\sim 5.0\times 10^{12}~\msun$ \citep{2008MNRAS.388..384D}. This makes this object one of the most massive lens galaxies ever observed. Follow-up spectroscopic analysis revealed the redshift of the lens and the source to be $0.444$ and $2.381$, respectively \citep{2009MNRAS.398.1263Q}.

\begin{figure*}%
\centering%
\includegraphics[width=.55\textwidth]{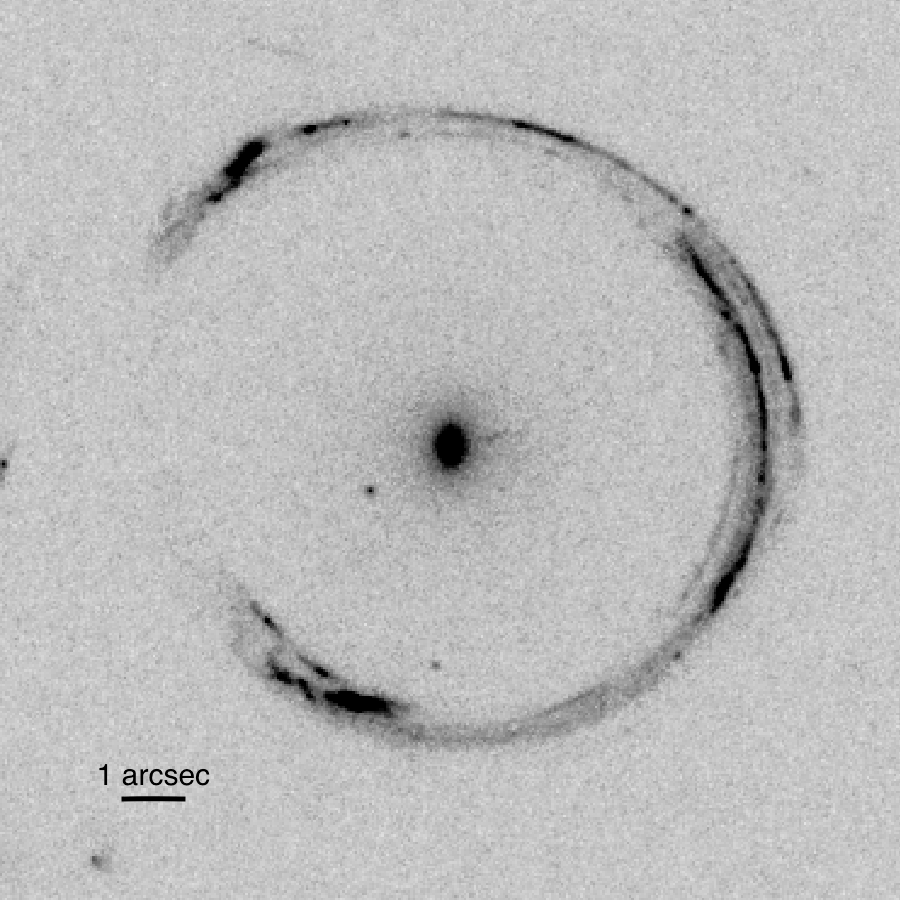}%
\\[\baselineskip]%
\setlength{\tabcolsep}{0pt}%
\renewcommand{\arraystretch}{1.5}%
\begin{tabular}{ccc}%
\includegraphics[width=.28875\textwidth]{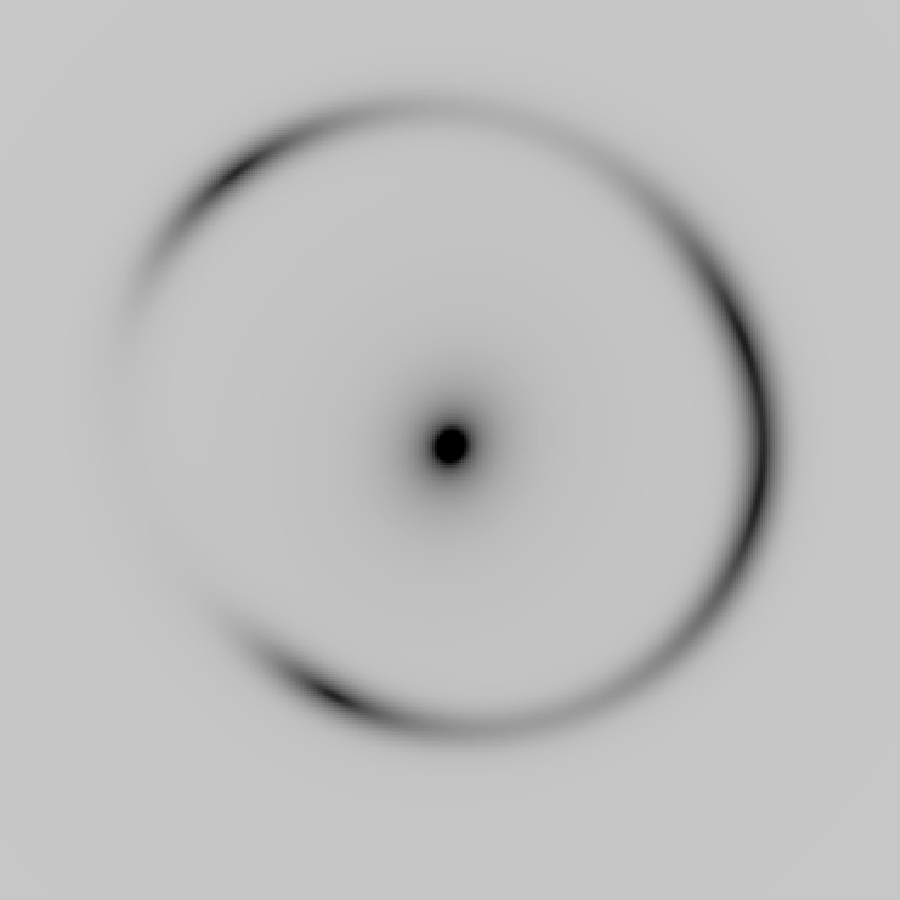}&%
\includegraphics[width=.28875\textwidth]{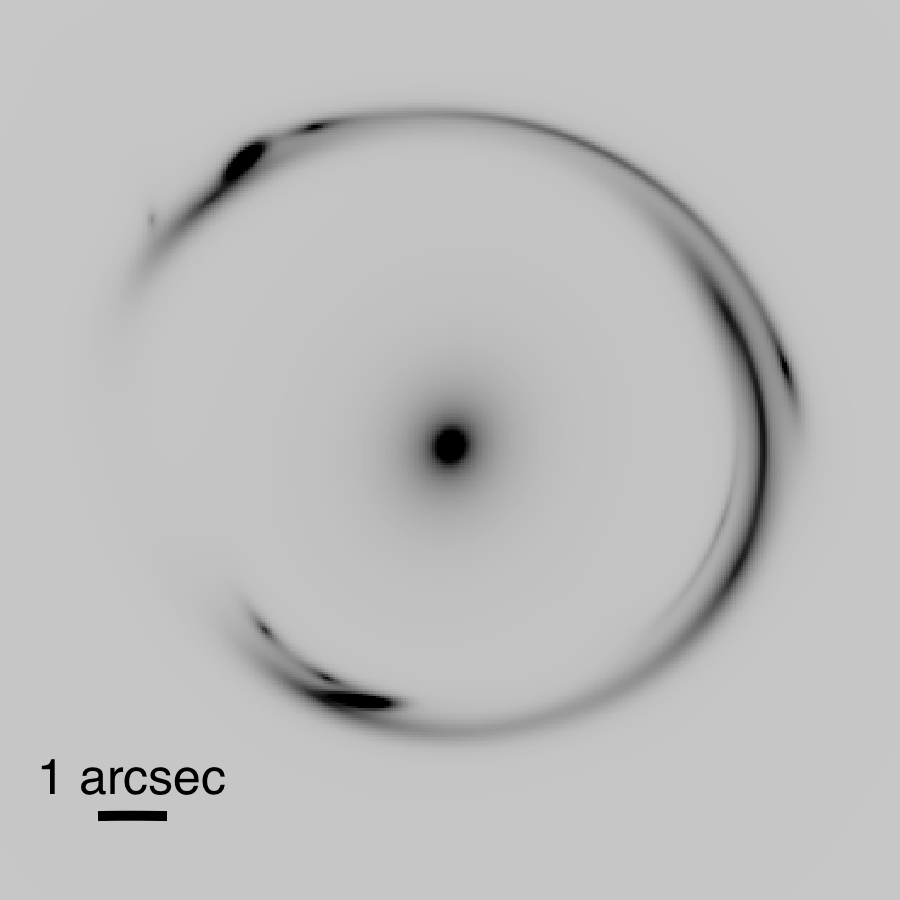}&%
\includegraphics[width=.28875\textwidth]{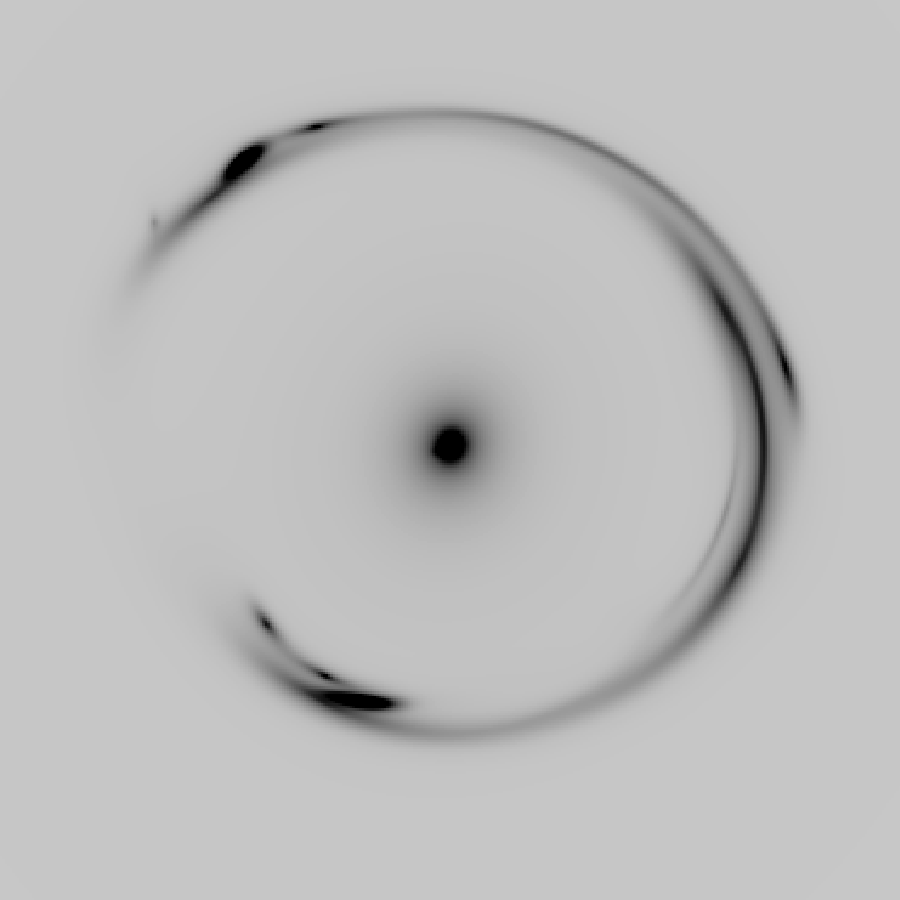}\\%
\includegraphics[width=.33\textwidth]{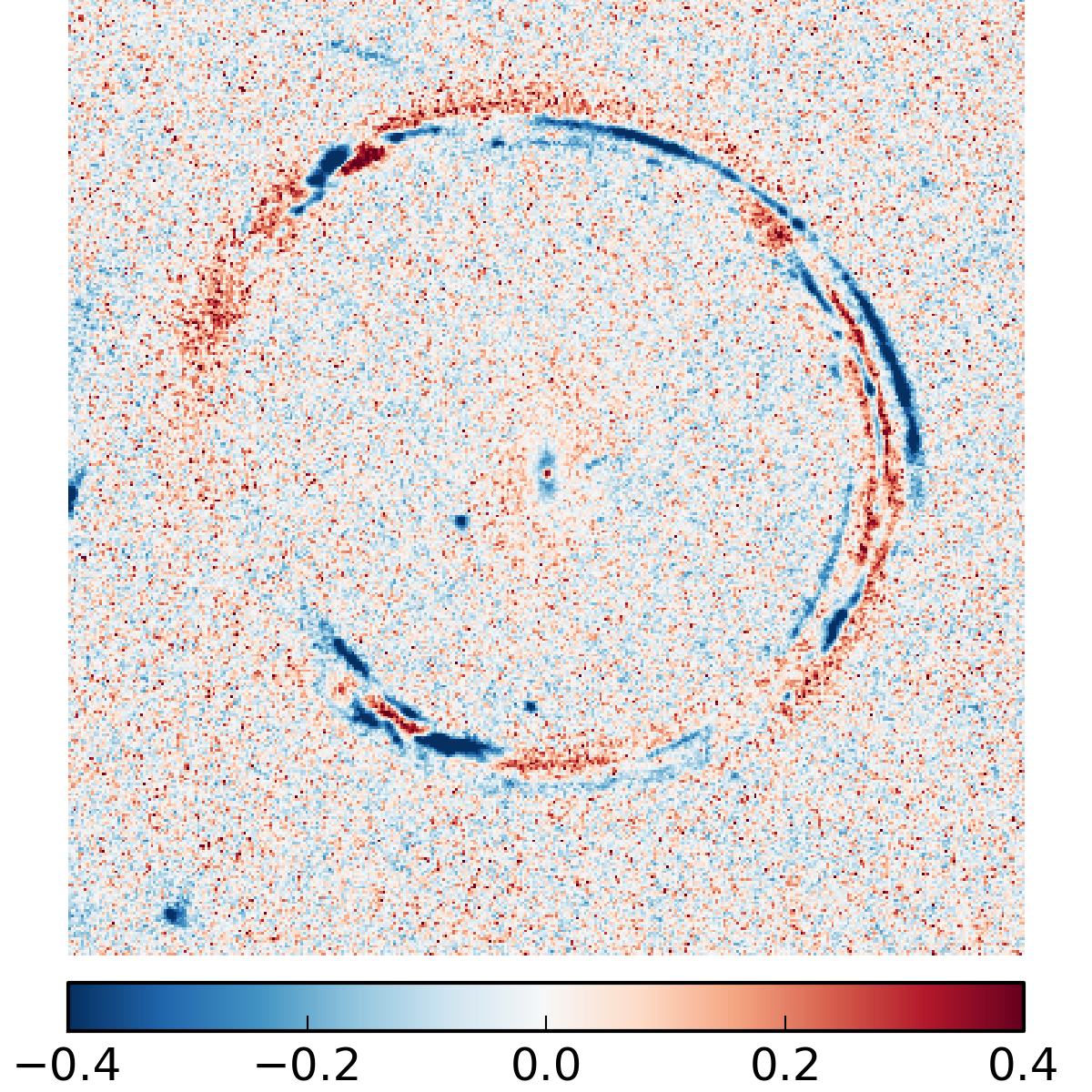}&%
\includegraphics[width=.33\textwidth]{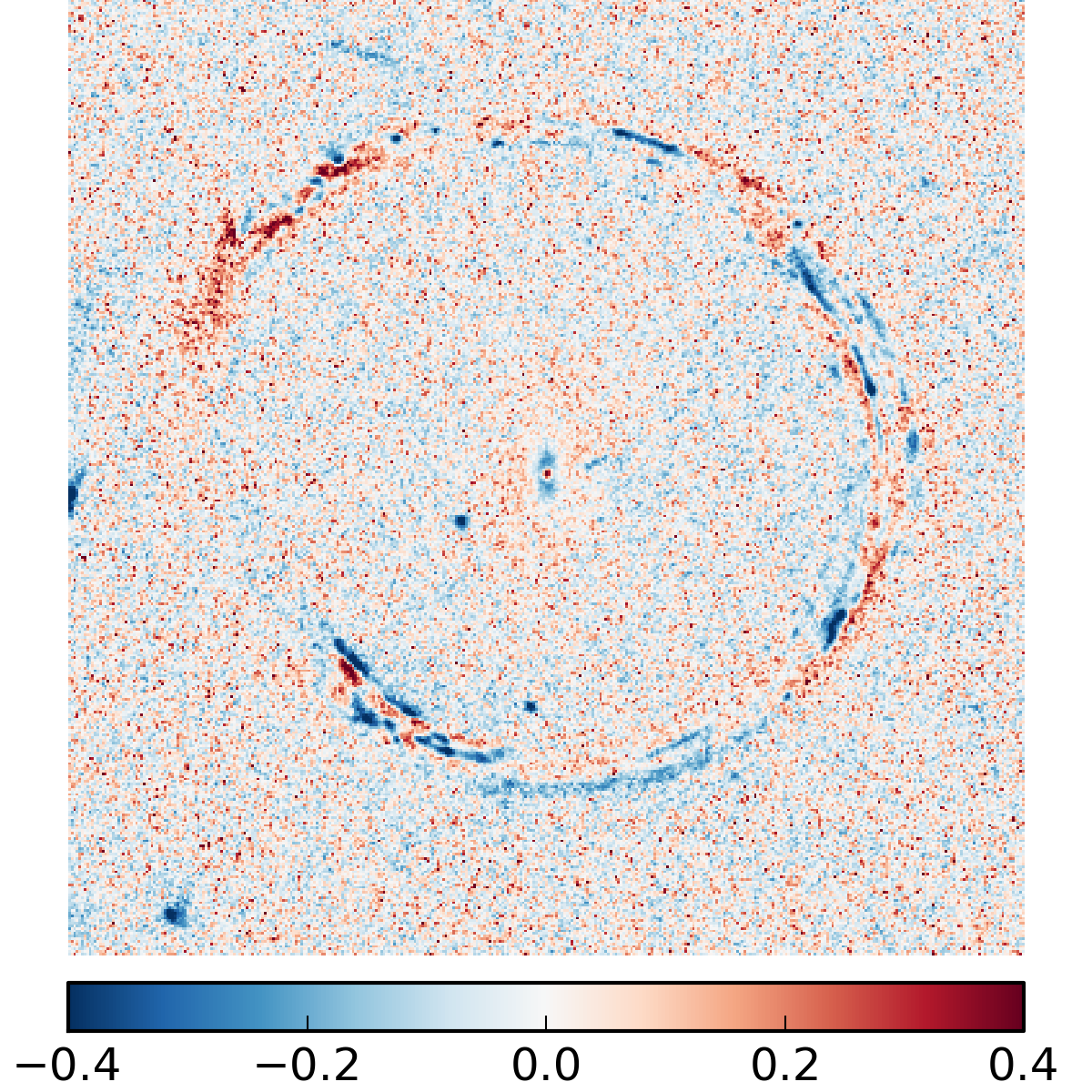}&%
\includegraphics[width=.33\textwidth]{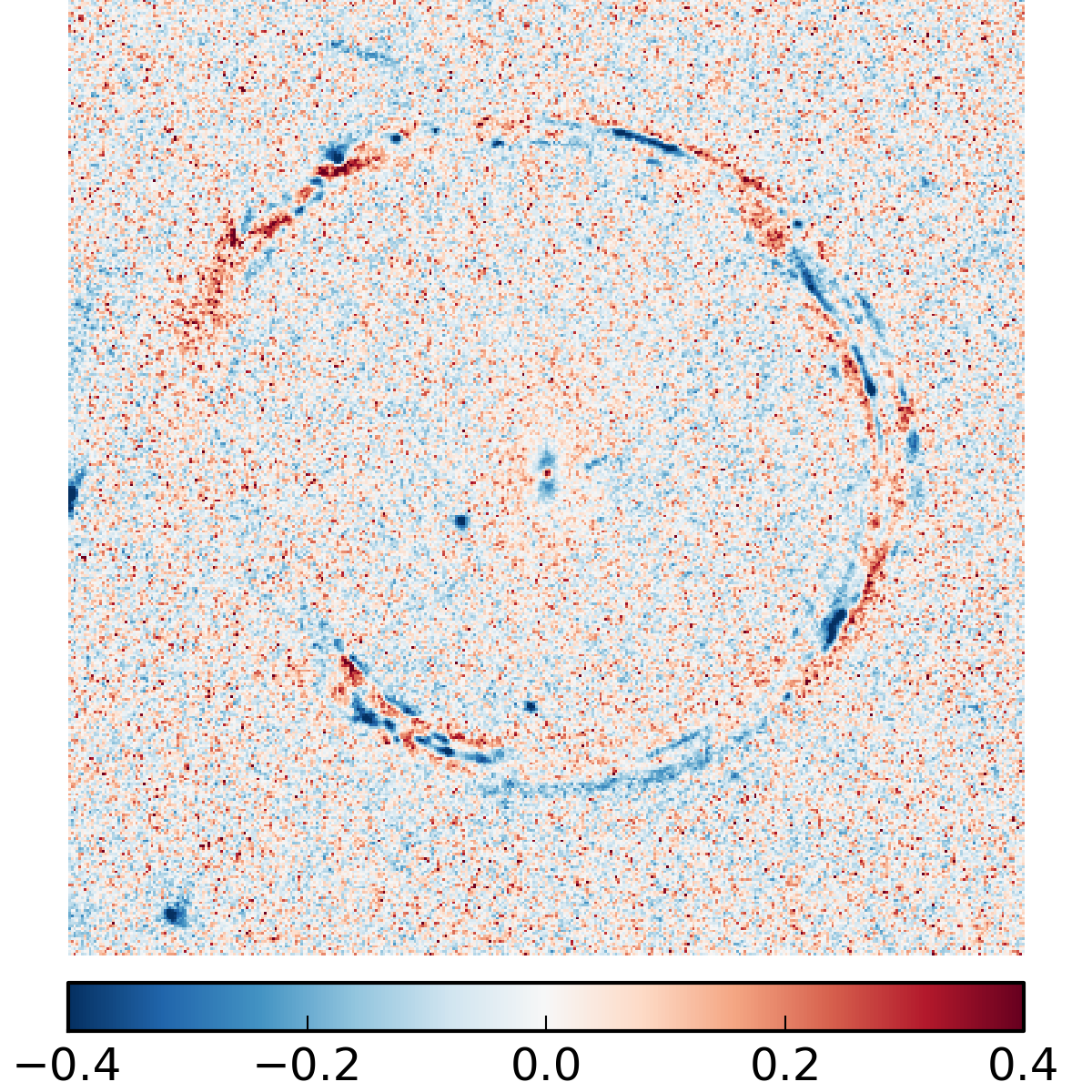}%
\end{tabular}%
\\%
\caption{%
\emph{Top panel:} Image of the Cosmic Horseshoe in the F475W band, with north up and east left.
\emph{Bottom panel:} Maximum-likelihood model images and residuals for three models presented in this paper: SIE with 1 source (\emph{left}), SIE with 5 sources (\emph{centre}), EPL with 5 sources (\emph{right}) . The residuals show the relative error (model-data)/data.
}%
\label{fig:image_and_models}%
\end{figure*}

The data we analyse in this work is taken from \textit{HST} observations with the Wide Field Camera under proposal 11602 and is freely available from the Hubble Legacy Archive \footnote{\url{http://hla.stsci.edu}}.
In particular, we use the image in the F475W band, which is the one with the best trade-off between angular resolution (0.04 arcsec after drizzling) and separation of the background object from the lens galaxy emission.
The resulting image, which is shown in the top panel of Fig.~\ref{fig:image_and_models} is a combination of 6 exposures, for a total observing time equal to 5454 secs.
We used Tiny Tim \footnote{\url{http://tinytim.stsci.edu}} \citep{2011SPIE.8127E..0JK} to obtain a model for the point-spread function.

%\newpage

\section{Method} \label{sec:method}
\subsection{Algorithm} \label{sec:algorithm}
The lens modelling analysis has been carried out with the \textsc{Lensed} code, described in detail in \citet*{2015arXiv150507674T}.
It is a new implementation of the forward reconstruction method, which takes advantage of a massively parallel ray-tracing kernel to perform the necessary calculations on a modern graphics processing unit (GPU) where available, or a traditional CPU.
We summarise the algorithm here.

Given an image to be analysed and a parametric model of choice, \textsc{Lensed} makes use of the Nested Sampling algorithm implemented in \textsc{MultiNest} \citep{2008MNRAS.384..449F,2009MNRAS.398.1601F,2013arXiv1306.2144F} to explore the parameter space of the model for user-defined priors.
For each set of parameters, the image is simulated by numerical integration of the deflected surface brightness distribution of the background sources.
Foreground objects such as the lens galaxy can be simulated as well and their parameters simultaneously recovered.
The model image can additionally be smoothed with the PSF of the observation.
Then the likelihood $P$ of the data $d$ given the model parameters $\xi$ is computed by comparing the real and model image on a pixel-by-pixel basis through the usual $\chi^2$ term

\begin{equation}\label{eq:logP}
	\log P(d \mid \xi) =
	-\frac{1}{2}  \sum_{i=1}^{N} \frac{(d_i - m_i)^2}{s_i^2} \;,
\end{equation}
where the sum runs over the $N$ pixels of the image, $m_i$ is the $i$-th pixel of the mock image and $s_i^2$ is the variance of the data for the same pixel. The posterior distribution of the parameters is then computed by use of Bayes' theorem
\begin{equation}\label{eq:posterior}
	P(\xi \mid d) = \frac{P(d \mid \xi) \, P(\xi)}{P(d)} \;,
\end{equation}
where $P(\xi)$ is the chosen prior probability of the set of parameters, which can take into account any previous knowledge about their distribution, e.g. luminosity function and ellipticity distribution for the background sources, or mass-to-light ratio for a given kind of galaxy.
The constant of normalisation $P(d)$ is the Bayesian evidence
\begin{equation}\label{eq:evidence}
	P(d) = \int P(d \mid \xi) \, P(\xi) d \xi\;.
\end{equation}

The scheme described above is straightforward and broadly applicable.
\textsc{Lensed}'s use of GPU computation and the efficient sampling technique make it robust and virtually free of parameter tuning by the user.
Here we list some of the features which are particularly relevant for the analysis presented in this paper:
\begin{itemize}
\item The parameter space is fully explored without any initial guess from the user except the selection of the priors.
As the posterior might have multiple modes, this capability prevents biases in the results arising from user prejudice.
\item The output contains the full posterior probability distribution, not only the set of maximum-likelihood parameters.
This allows us to analyse potentially multiple solutions and correlations between the parameters.
\item The code is well-suited for as many as ${\sim}~60$ parameters.
This turns out to be necessary in order to describe the source and lens configuration of the complicated observation considered here with reasonable precision.
\item Due to the speed of GPU calculation, it is very fast as well as very accurate in simulating the model images.
This allows us to work with the original image at full resolution, and to sample the parameter space at the necessary density.
\item The model configuration is very flexible.
We can build the desired model starting from a simple one and subsequently add components to the lens and the source as we better understand the nature of the objects.
\end{itemize}

\subsection{Lens models} \label{sect:lens_models}
In our reconstructions, the mass distribution of the lens is described by a number of different analytical models.
The simplest lens model we consider is the singular isothermal ellipsoid (SIE), for which the dimensionless surface mass density (convergence)~$\kappa$ is given by 
\begin{equation}\label{eq:sie}
\kappa(R) = \frac 1 2 \frac b R ,
\end{equation}
where $b$ is a scale radius and $R$ is the elliptical radius defined by
\begin{equation}
R = \sqrt{x^2q + y^2/q} ,
\end{equation}
in coordinates where the major axis and minor axis of the ellipse coincide with the $x$-axis and $y$-axis of the coordinate system, respectively, and $0 < q < 1$ is the axis ratio. 
In the spherical limit $q = 1$, the parameter $b$ is the Einstein radius of the lens.
See \citet{1994A&A...284..285K} for a detailed analysis of the lensing properties of this model.

We also use an elliptical power law (EPL) lens model, which is generalisation of the SIE model where the radial slope $t$ of the mass distribution is an additional parameter.
It has the surface density
\begin{equation}\label{eq:epl}
\kappa(R) = \frac {2-t} {2} \left( \frac b R \right)^t ,
\end{equation}
where the slope $t$ falls into the range $0 < t <2$, and the case~$t = 1$ recovers the SIE profile.
The normalisation is chosen so that $b$ retains its meaning as the approximate Einstein radius of the lens, in the sense that an ellipse with elliptical radius~$R = b$ contains mass~$\pi \, b^2$, independent of the choice of~$q$ and~$t$.
Unlike a SIE, this model does not admit an analytical expression for the deflection angle.
Instead, we follow the approach of \citet{2015A&A...580A..79T} for the efficient numerical calculation of the required quantities.

When we want to account for the mass distribution of the stellar component, we employ the \citet{1990ApJ...356..359H} model, with surface mass density
\begin{equation}
\kappa(x) = \frac {\kappa_s} {(x^2-1)^2}  [(2+x^2)f(x)-3] ,
\end{equation}
where
\begin{equation}
f(x) = \left\{ \begin{array}{cl}
\displaystyle \frac { \arctan{\sqrt{x^2-1}}} { \sqrt{x^2-1} } &  \textrm{if } x > 1 \\
\\
\displaystyle \frac { \arctanh{\sqrt{1-x^2}}} { \sqrt{1-x^2} }& \textrm{if } x < 1 \\
\\
1 & \textrm{if } x = 1
\end{array} \right. 
\end{equation}
and $x$ is the (circular) radius in units of the scale radius~$r_s$.
This mass distribution is commonly used to approximate the 3-d properties of the light distribution of \citet{1948AnAp...11..247D}, which is typical of early-type galaxies.
 
We further add to the lens models an external shear component $\gamma = (\gamma_1, \gamma_2)$, which accounts for shear contribution from the mass distribution outside the main lens. The shear component $\gamma_1$ is aligned with the x-axis of the image, while $\gamma_2$ is rotated by $45^\circ$ counter-clockwise.

\subsection{Source models} \label{sect:source_models}
Ideally, in a parametric method, one would like to use a simple yet realistic and complete model for the background source.
Simplicity is desired because as the number of model parameters increases, the problem becomes more and more computationally difficult.
While the flux distribution of some types of galaxies (e.g. early-type ellipticals) can be described analytically with good precision, many galaxies at high redshift exhibit an irregular morphology which precludes a simple representation.
Moreover, the common observation of interacting and merging galaxies makes it difficult to discern unambiguously between a single object with complex morphology and a close group of possibly interacting galaxies.
This fact becomes even more problematic when dealing with highly gravitationally magnified images where the multi-component morphology of the background source is clearly visible.

In order to overcome this problem and, at the same time, not be limited in the lens reconstruction by an unsatisfactory modelling of the source, we start from a simple one-component S\'{e}rsic profile, and we subsequently increase the number of S\'{e}rsic components until two conditions are met: the lens parameters no longer change significantly, and there are no image components clearly unresolved by our reconstruction.
Although the source configuration resulting from this procedure may not be physically motivated, we achieve a reasonably elastic description of the source brightness distribution, as each new component can reproduce one of the blobs observed in the multiple images.
At the same time, the fact that we are using a parametric model allows us to measure directly the properties of each single component, as well as derive some global quantities, such as e.g. the total flux.

We also model the lens galaxy with a S\'{e}rsic profile, leaving the center as a free parameter, i.e. we do not force the mass and the light to be perfectly aligned.
In many galaxy-galaxy strong lenses, accurate modelling of the central galaxy is required, because its light overlaps significantly with the lensed background images.
In contrast to other methods which subtract the host galaxy in a separate preliminary step \citep{2006ApJ...638..703B, 2007ApJ...671.1196M}, \textsc{Lensed} optimises the parameters of the host galaxy light at the same time as the lens mass and the background light.
We note that in the case of the object analysed in this paper, this is not a critical task, because the lensed images appear at positions where the light coming from the lens galaxy is negligible, at least in the band we have chosen for our analysis.

\section{Results} \label{sec:results}

\begin{figure}
 \includegraphics[width=\columnwidth]{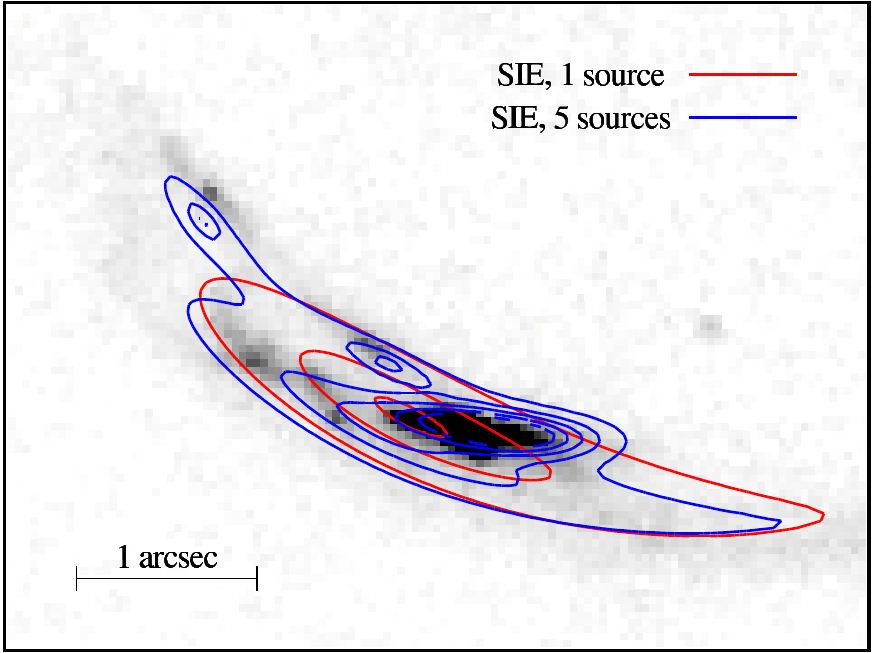}
 \caption{Zoom on the bottom-left image of the system shown in Fig.~\ref{fig:image_and_models}.
The contours show the brightness distribution of the maximum-likelihood model images for the SIE model using 1 background source component (\emph{red}) and 5 background source components (\emph{blue}).}
 \label{fig:sie_image_cont}
\end{figure}

\begin{figure*}
 \includegraphics[width=\columnwidth]{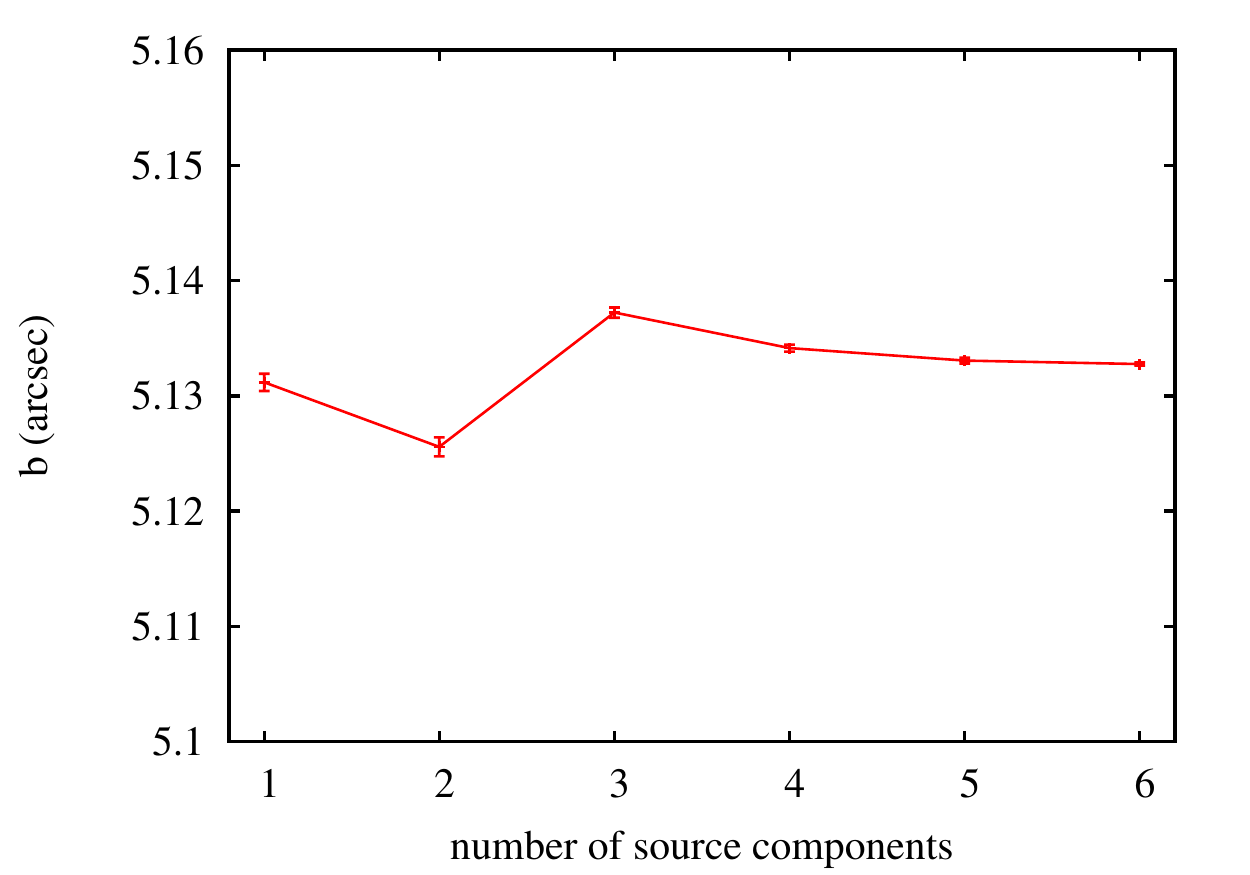}
 \includegraphics[width=\columnwidth]{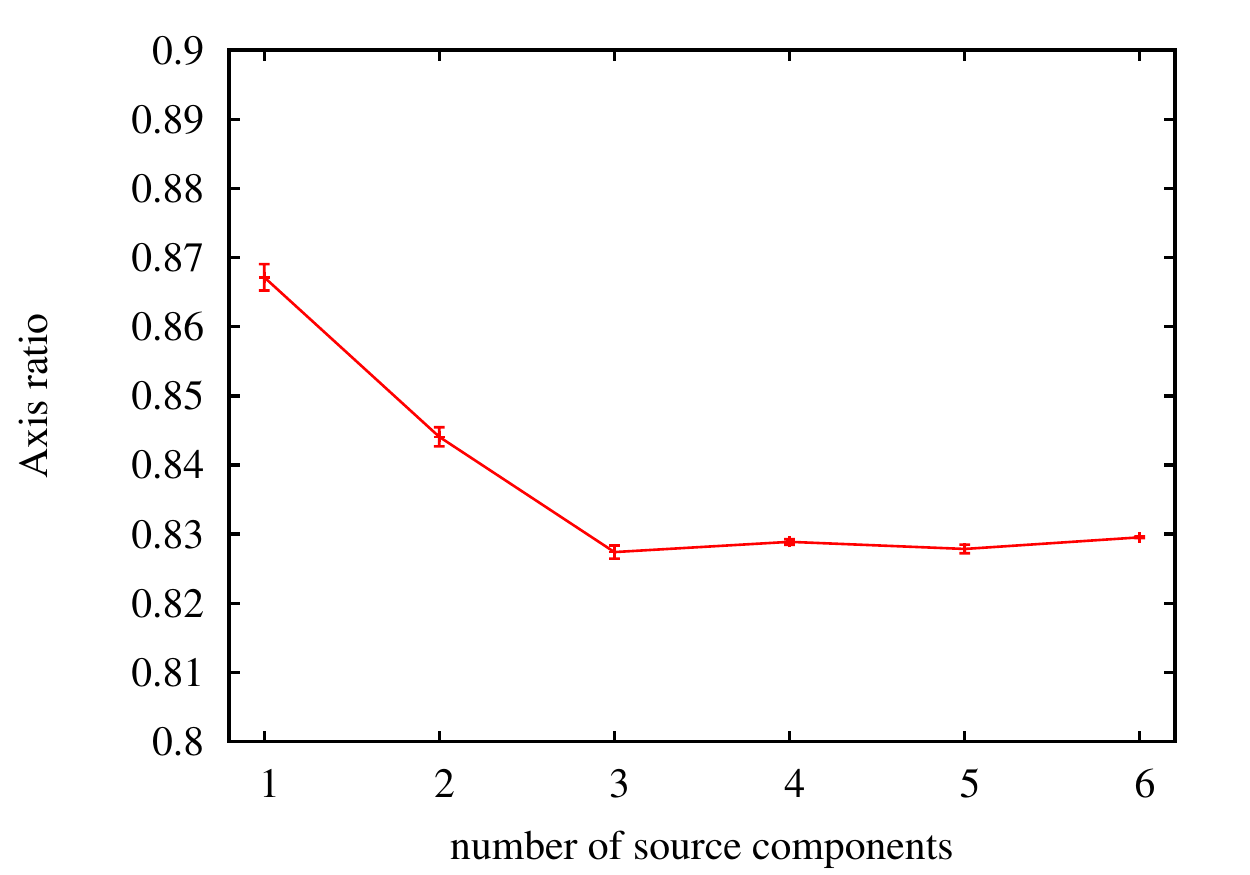}
 \includegraphics[width=\columnwidth]{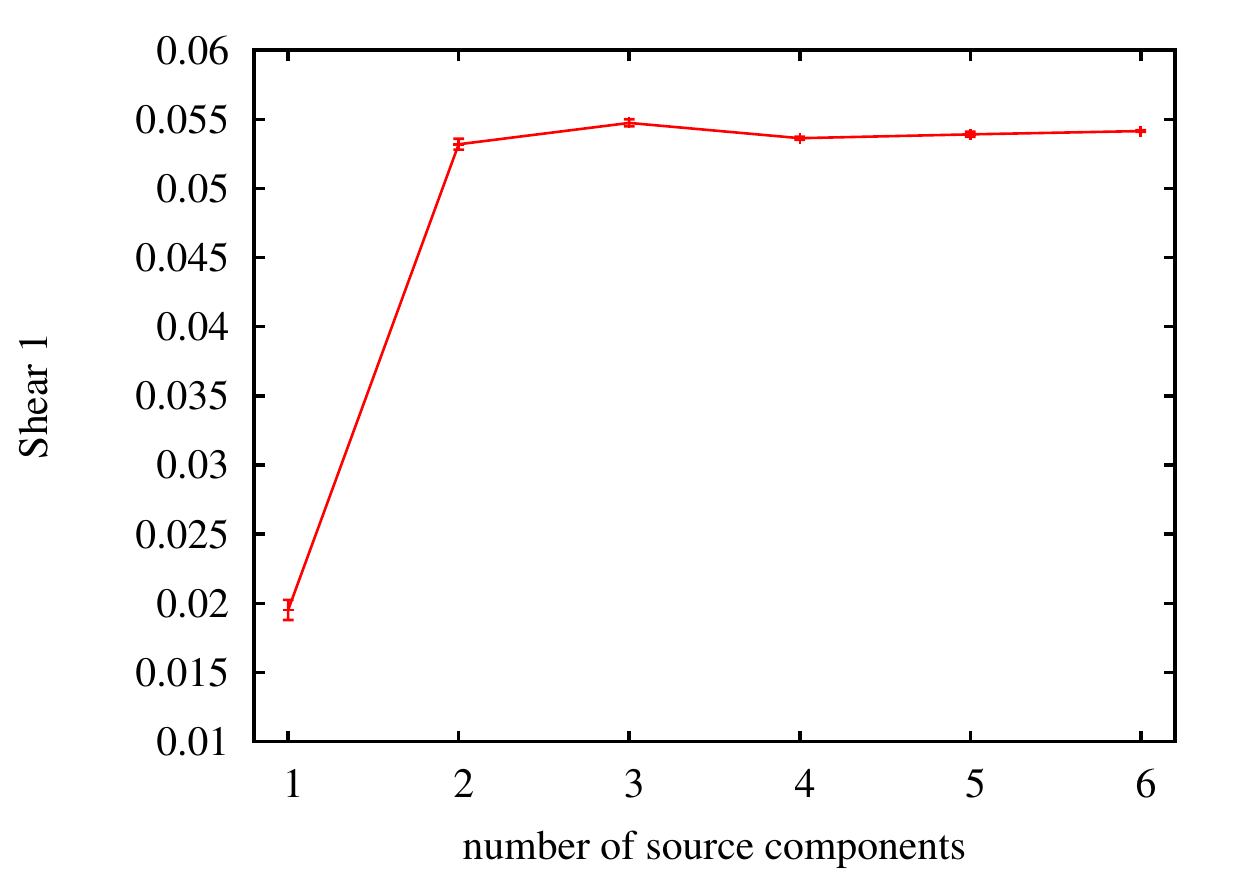}
 \includegraphics[width=\columnwidth]{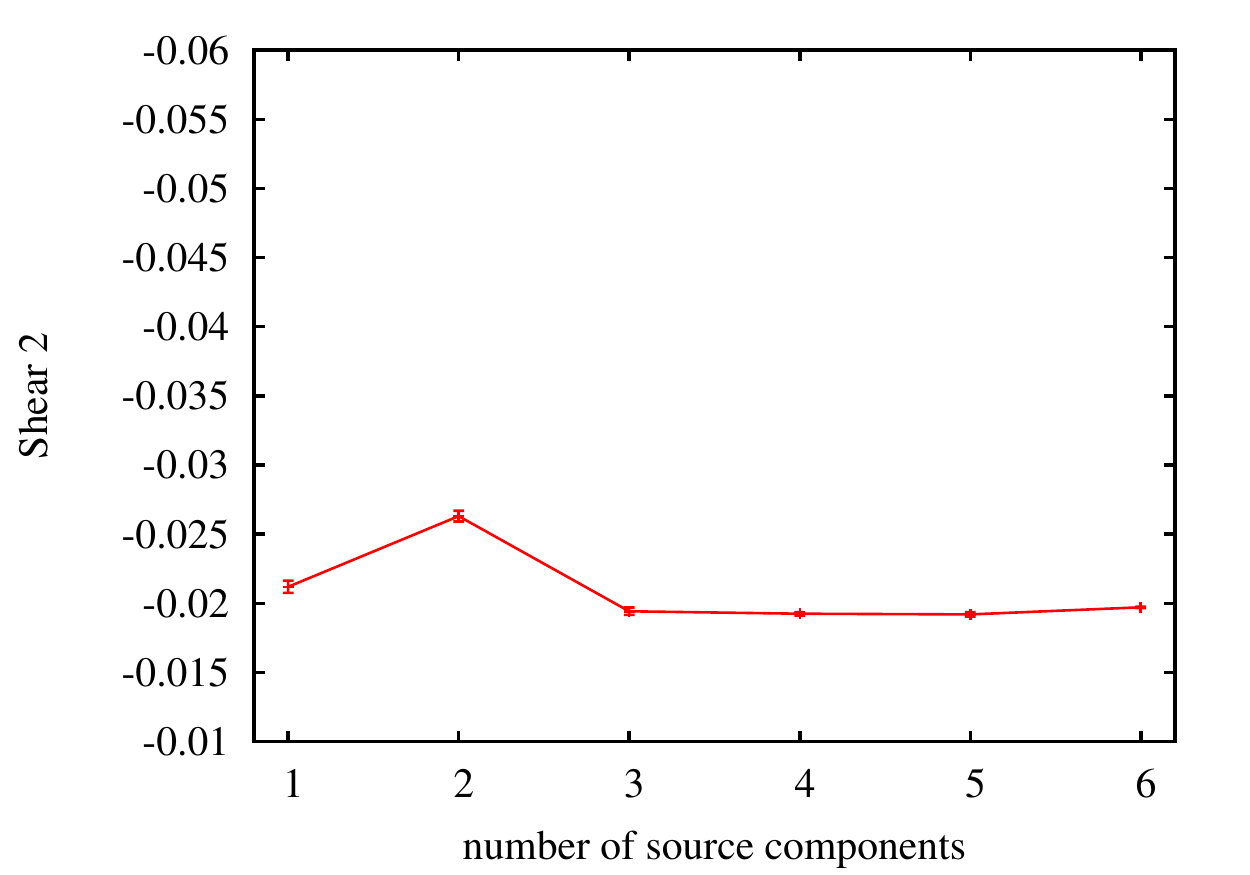}
 \caption{Different lens parameters for the SIE+$\gamma$ model as a function of the number of background sources.}
 \label{fig:sie_num_compo}
\end{figure*}

\subsection{Determining the number of source components}\label{sect:num_compo}
We apply the procedure described in Section~\ref{sect:source_models} to the simplest lens model we consider, the singular isothermal ellipsoid with external shear.
In the left bottom panel of Fig.~\ref{fig:image_and_models} we show the resulting image and residuals for the model with one S\'{e}rsic source component.
We note that even with this simple model, we recover the global shape of the ring, although the complex details of the arc are not reproduced.
This smooth model visually resembles the ground-based observations at lower resolution shown by \citet{2007ApJ...671L...9B} and \citet{2008MNRAS.388..384D}, and a single source component would likely be sufficient to reconstruct these images.

In the \textit{HST} observations we are considering, many sub-images are clearly visible, and a smooth single-component model cannot be considered satisfactory, because the visible sub-structure of the arcs presents additional information to reconstruct the lens parameters.
For this reason, we subsequently increase the number of background source components until all the major features of the image are reproduced and the lens parameters do not change significantly anymore. 
In Fig.~\ref{fig:sie_image_cont} one of the images is highlighted and it is shown how the surface brightness contours change when additional sources are added.
It is clearly visible that the different components are optimised to recover the visible sub-structure in the image, as far as compatible with the limitations of the simple lens model.

For the rest of the paper, we will present results for 5 background source components, because this is the minimum number required to reproduce the main features of the arcs. This can be seen in the central bottom panel of Fig.~\ref{fig:image_and_models}, where we show the reconstructed image with an SIE lens and  5 sources.
Comparing it with the top panel of the same figure, we see that all the major components of the image are reproduced.
As shown in Fig.~\ref{fig:sie_num_compo}, the lens parameters tend to converge as the number of sources increases and are practically stable for $\ge$ 4 source components. We checked that the same argument is valid for all the lens models presented in the following sections.

\begin{table*}
	\centering
	\caption{Lens parameters for different slopes of the mass profile. The SIE case is $t=1$.  All models were constructed with 5 S\'{e}rsic sources. The position angle (P.A.) indicates the direction of the major axis of the elliptical lens, measured east of north. }
	\label{tab:slope_lens_param}
	\setlength{\tabcolsep}{.9\tabcolsep}
	\begin{tabular}{cccccccc}
		\hline
		prior on $t$ & $t$ & $b$ (arcsec) & $q$ & P.A. & $\gamma_1$ & $\gamma_2$ & $\log P(d \mid \xi)$\\
		\hline
0 < t < 2 & $0.0821 \pm 0.0017$ & $5.1563 \pm 0.0003$ & $0.8901 \pm 0.0004$ & $24.4 \pm 0.1$ & $0.0384 \pm 0.0002$ & $ 0.0376 \pm 0.0002$   & $-147\,690$ \\
fixed     & $0.5$               & $5.1426 \pm 0.0003$ & $0.8742 \pm 0.0006$ & $24.3 \pm 0.1$ & $0.0438 \pm 0.0003$ & $ 0.0107 \pm 0.0002$   & $-148\,426$ \\
fixed     & $1.0$               & $5.1331 \pm 0.0003$ & $0.8279 \pm 0.0006$ & $23.4 \pm 0.1$ & $0.0539 \pm 0.0002$ & $-0.0192 \pm 0.0002$~~ & $-149\,556$ \\
fixed     & $1.5$               & $5.1213 \pm 0.0002$ & $0.7394 \pm 0.0008$ & $22.5 \pm 0.1$ & $0.0587 \pm 0.0001$ & $-0.0528 \pm 0.0001$~~ & $-150\,915$ \\
		\hline
	\end{tabular}
\end{table*}

\subsection{Constraining the slope of the mass distribution}\label{sect:constr_slope}

\begin{figure*}%
\centering%
\begin{minipage}[b]{0.5\linewidth}%
\includegraphics{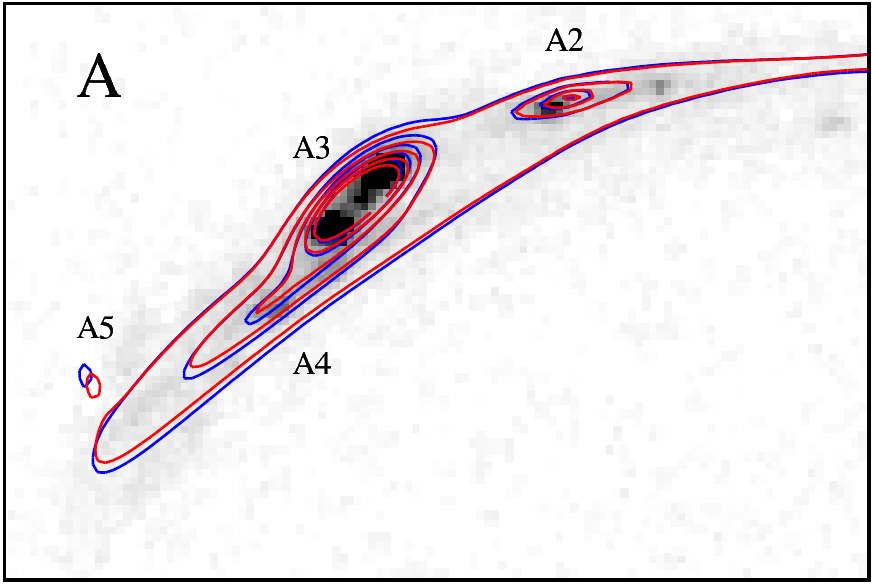}%
\\[1.37cm]%
\includegraphics{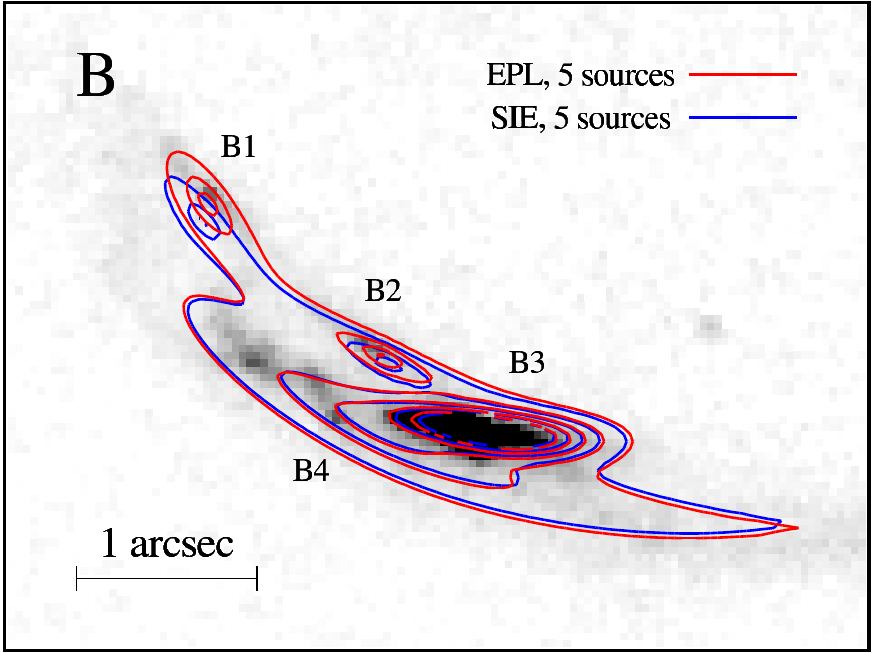}%
\vspace{1.5pt}%
\end{minipage}%
\hspace{.5cm}%
\begin{minipage}[b]{0.31\linewidth}%
% the scale is needed to get the image with the same pixel size of the other two
\includegraphics[scale=1.5833]{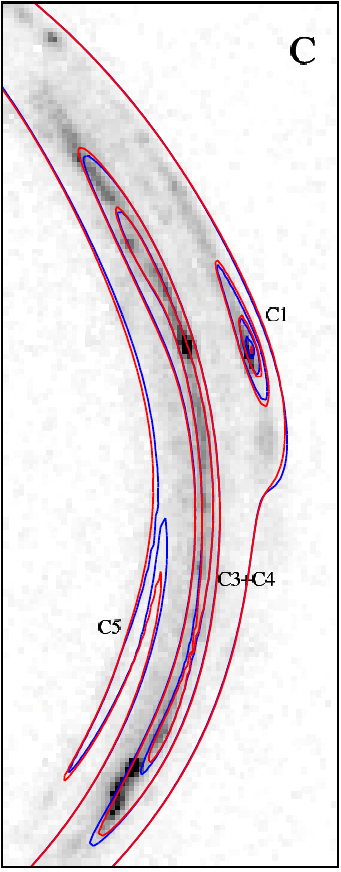}%
\end{minipage}%
\caption{%
Zoom on the three images of the system: A on the top left, B on the bottom left, C on the right.
The contours show the brightness distribution of the images for the maximum-likelihood EPL (\emph{red}) and SIE (\emph{blue}) models with 5 source components.
The numbers refer to the source components shown in Fig.~\ref{fig:sources_caustic}.
}%
\label{fig:slope_image_cont}%
\end{figure*}

We use the EPL model described in Section~\ref{sect:lens_models}.
With respect to the results presented in Section~\ref{sect:num_compo}, there is an additional free parameter, which is the slope $t$ of the mass distribution.
The bottom-right panel of Fig.~\ref{fig:image_and_models} shows the maximum-likelihood EPL+$\gamma$ model.
Table~\ref{tab:slope_lens_param} shows a comparison of the lens parameters between EPL models with free parameter $t$ and a number of special cases with fixed values of $t$. Here and in the following, we do not quote the resulting values for the center of the lens, which is always well aligned with the centroid of the light distribution, with a typical offset between 2 and 4 pixels (0.08 to 0.16 arcsec).
From the likelihood values in Table~\ref{tab:slope_lens_param}, we see that the model becomes less likely as the slope~$t$ increases.
The preferred slope is extremely flat, with a mean posterior value of~$t = 0.082$.
In order to better understand the origin of this value, we show the comparison between the image distributions of the maximum-likelihood EPL model and the SIE model in Fig.~\ref{fig:slope_image_cont}.
The main difference between the two models is the position of the image B1.
In Section~\ref{sec:degeneracy}, we will investigate the physical origin of the preference for a flatter-than-isothermal slope.

\begin{figure*}%
\includegraphics[width=\columnwidth]{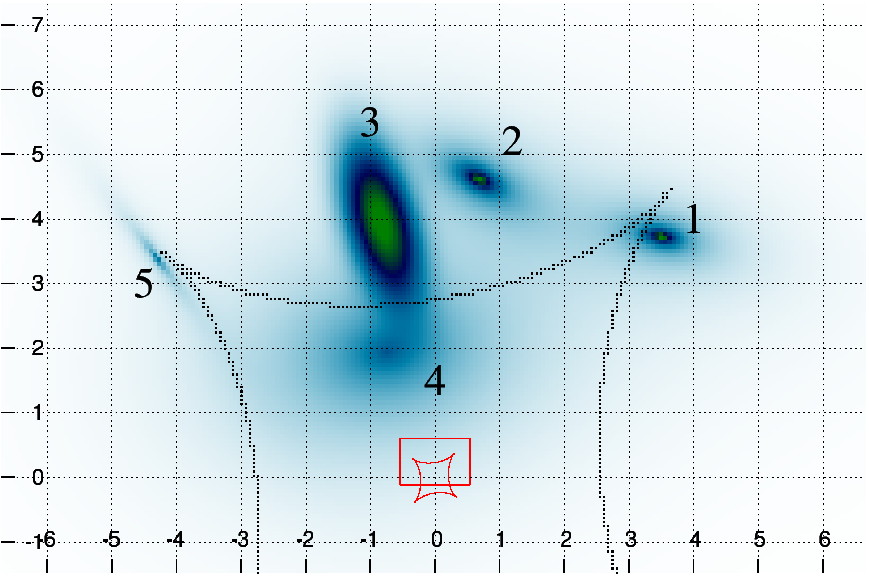}%
\hfill%
\includegraphics[width=\columnwidth]{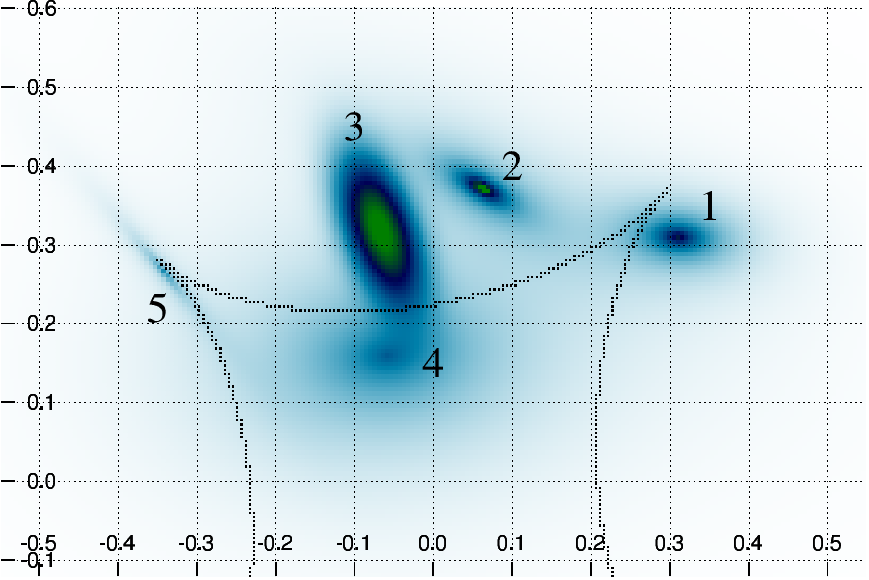}%
\caption{%
Sources and caustic for two different mass models: on the left the maximum-likelihood SIE+$\gamma$, and on the right the maximum-likelihood EPL+$\gamma$ with slope~$t = 0.082$. The numbers on the axis show the positions with respect to the lens center in kpc. The border and the caustics of the right panel are reproduced in red inside the left panel in order to highlight the different scale of the two figures.
}%
\label{fig:sources_caustic}%
\end{figure*}

\begin{figure*}
 \includegraphics[width=0.66\columnwidth]{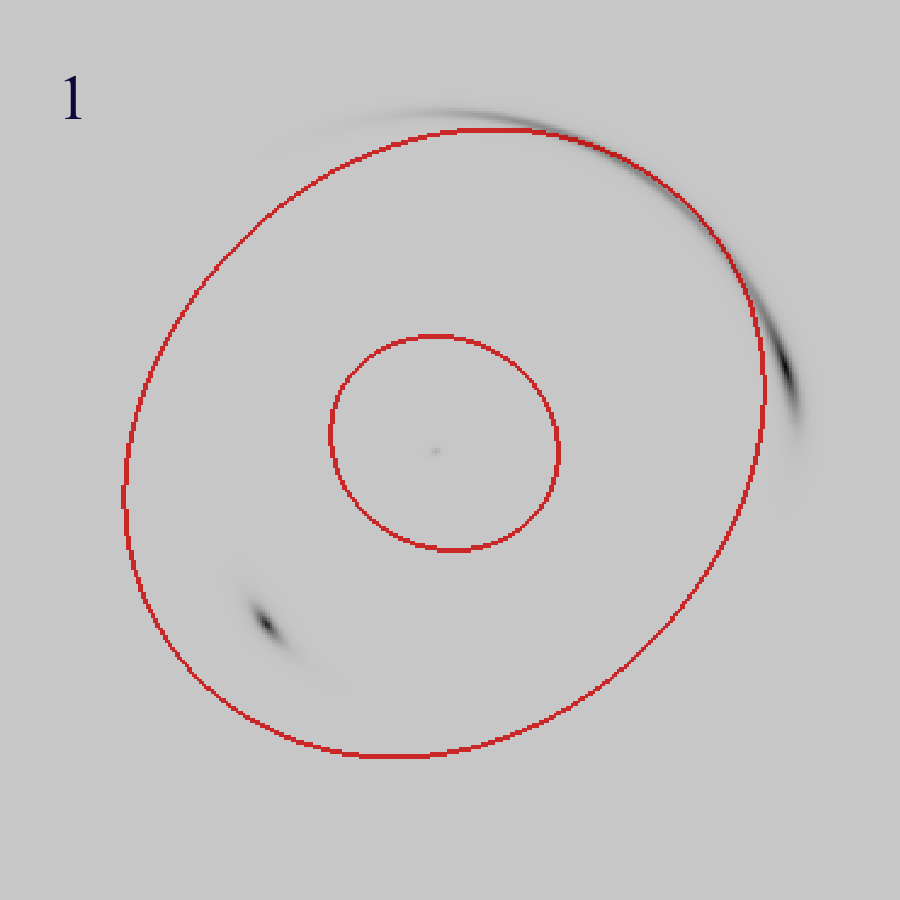}
 \includegraphics[width=0.66\columnwidth]{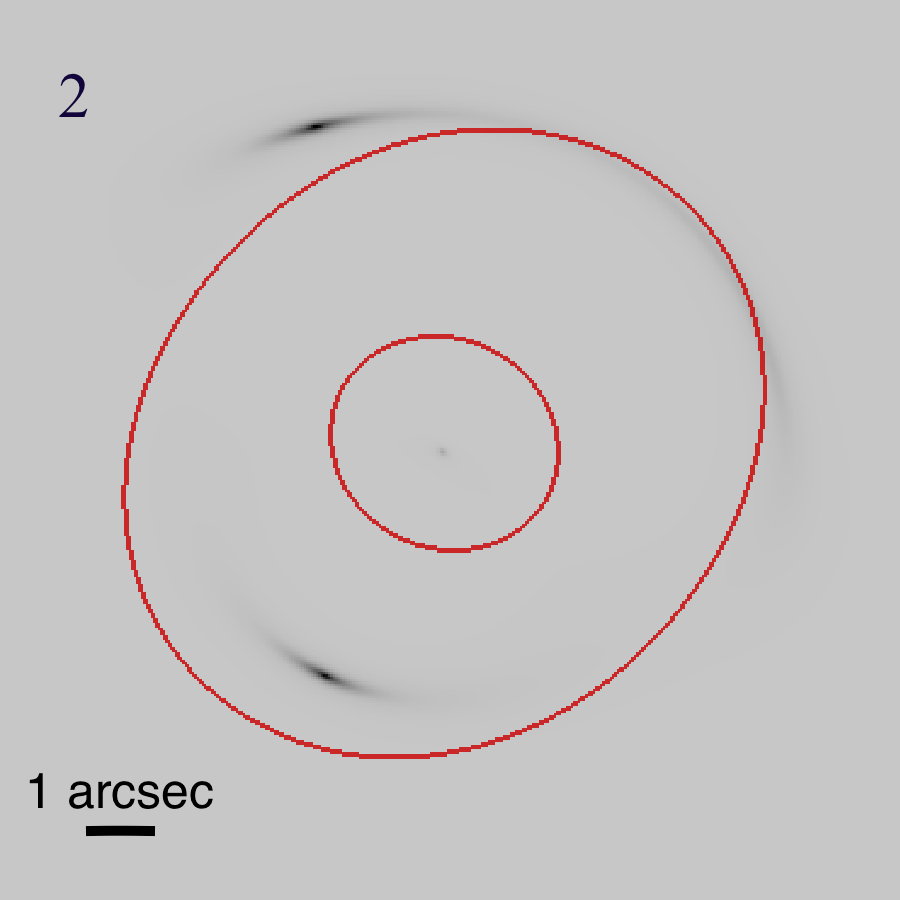}
 \includegraphics[width=0.66\columnwidth]{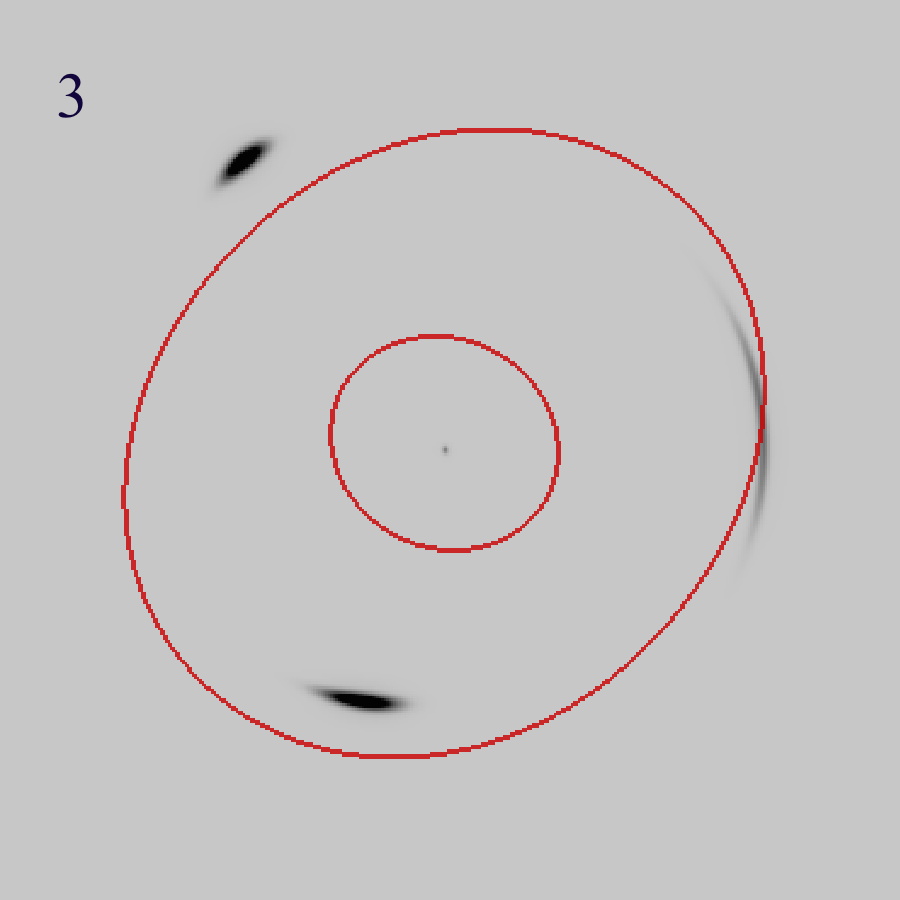}
 \includegraphics[width=0.66\columnwidth]{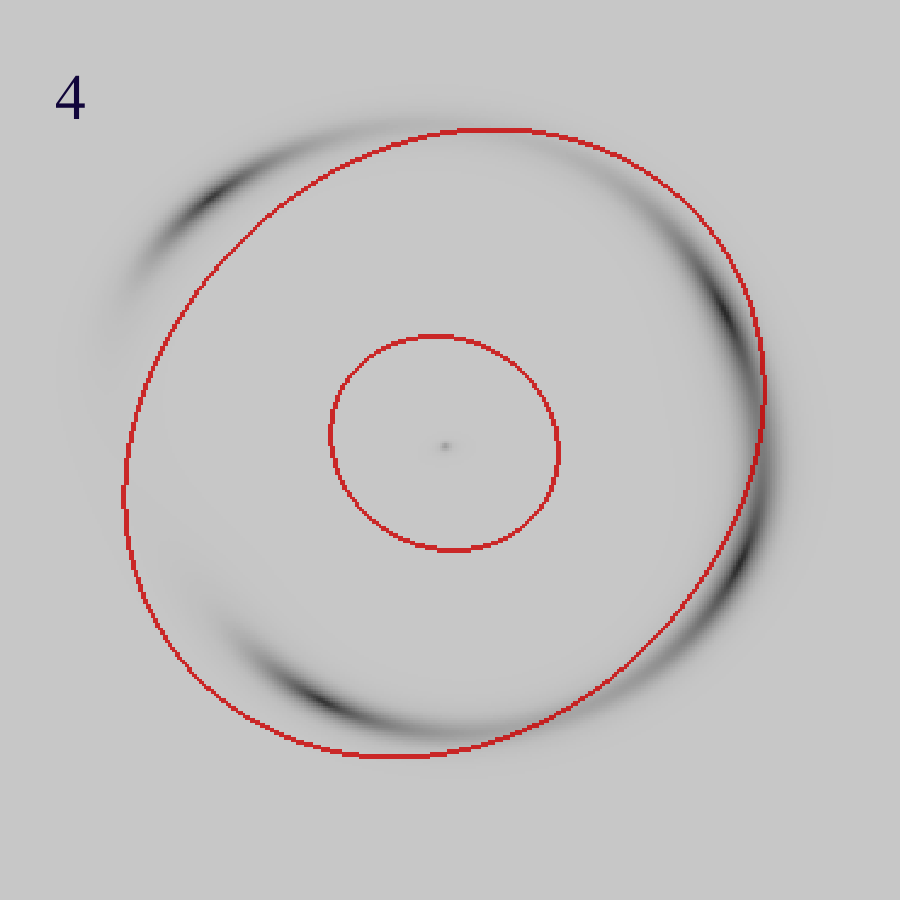}
 \includegraphics[width=0.66\columnwidth]{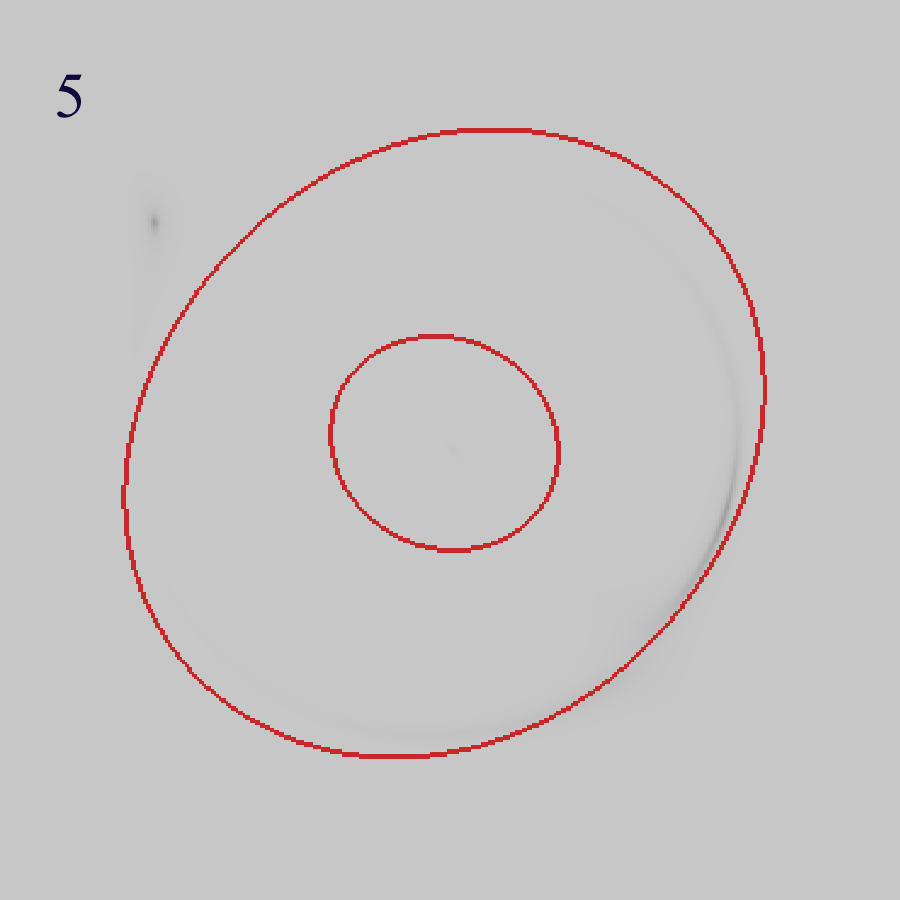}
 \caption{Images of the different source components for the maximum-likelihood EPL+$\gamma$ model. The corresponding appearence on the source plane can be seen in Fig.~(\ref{fig:sources_caustic}).
 The critical lines are marked in red. 
}
 \label{fig:epl_images}
\end{figure*}

In Fig.~\ref{fig:sources_caustic} we show the source plane configurations for the two models, and Fig.~\ref{fig:epl_images} shows the images that correspond to the different source components. 
For simplicity we use only the preferred EPL+$\gamma$ model, but the geometry of the images for other models is very similar, as one expects given the similar source positions relative to the caustics.
From the separate images, it is possible to appreciate how the apparent three-image configuration is actually made up of source components with different multiplicities on the image plane. Please note that the numbering of the sources is totally arbitrary, and it does not reflect any specific ranking.

The width of the posterior distribution of the EPL model obtained by the algorithm is very small, with a standard deviation for the slope $t$ of order~$\sigma_t \sim 10^{-3}$.
Since this is a statistical constraint derived under the assumption that the assumed model is in fact the true one, care must be taken when interpreting this result as the uncertainty on the slope of the real mass distribution, which almost certainly does not follow a perfectly elliptical power law profile.
In order to give a broader meaning to the parameters, i.e. by interpreting~$t$ as the mean slope of the lens mass distribution, it is necessary to add a model uncertainty to the error given by the algorithm.

However, the observable mismatch between the best SIE model and the image configuration, visible in the position of the B1 image in Fig.~\ref{fig:slope_image_cont}, makes us believe that the isothermal profile is excluded with a high degree of confidence. Also, this kind of positional mismatch cannot be attributed to any particular choice of ours in the source model or configuration. 

\subsection{Two-component mass model}\label{sec:two-compo}
\begin{figure}%
\includegraphics[width=\columnwidth]{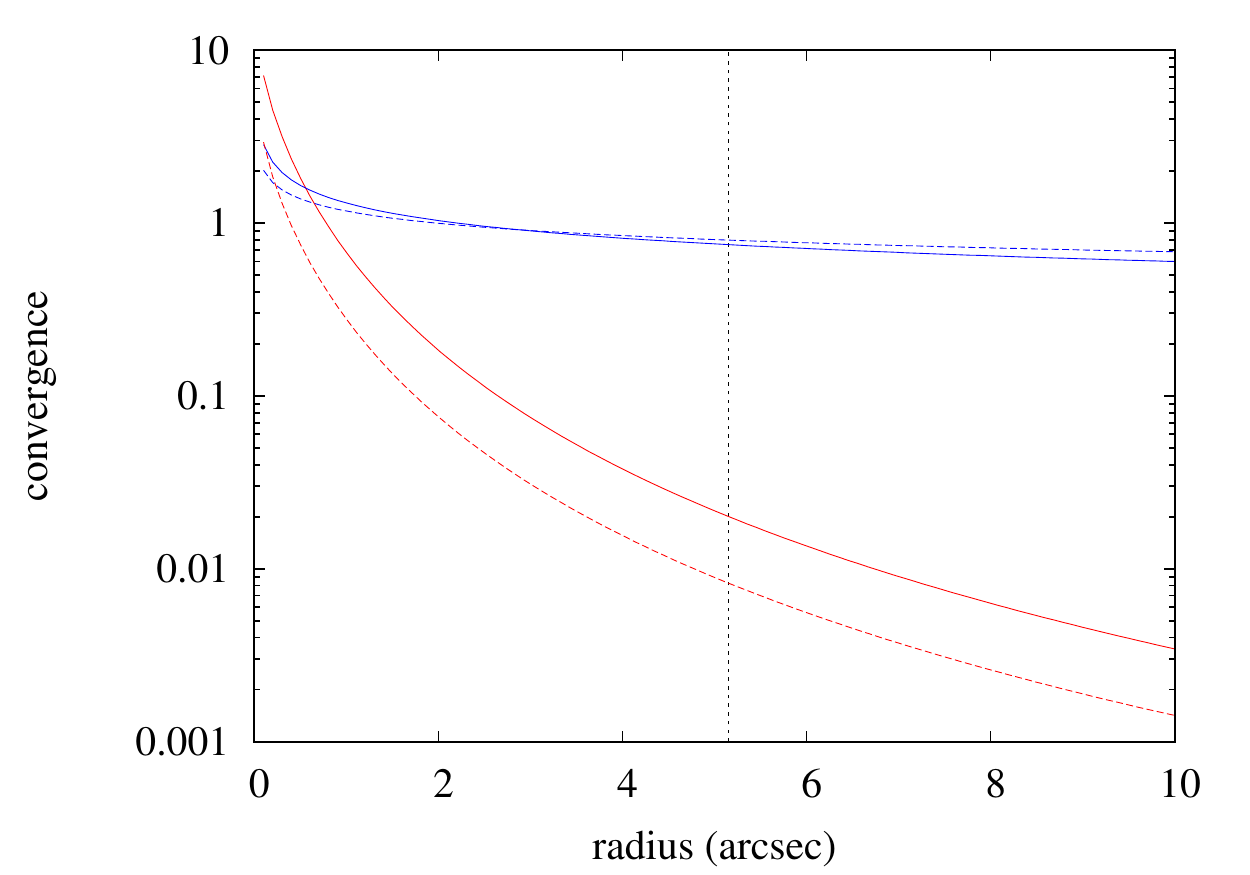}%
\caption{%
Profile of the projected mass density profile for the different mass components in units of the critical density $\Sigma_\text{crit}$.
The models with the Salpeter and Chabrier IMFs are marked with solid and dashed lines, respectively.
In each case, the blue line represents the dark matter component, and the red line represents the stellar component.
The vertical black line marks the Einstein radius of the object.
}%
\label{fig:lens_profile}%
\end{figure}

We now try and distinguish between the mass distribution of the baryonic component of the lens galaxy and the dark matter profile.
As detailed in Section~\ref{sect:lens_models}, we model the stars with a Hernquist distribution, a convenient approximation of the de Vaucouleurs profile for ray-tracing calculations \citep[see][]{1990ApJ...356..359H}.
This assumption is acceptable given the parameters we find for the lens light profile (S\'{e}rsic index $n \sim 4$).
The centre of the lens galaxy is fixed to the center of the light distribution, the scale radius $r_s$ is by construction set to $0.5509 \times R_\text{eff}$ of the S\'{e}rsic profile \citep{2006glsw.conf.....M}.
The only free parameter is then the convergence scale~$\kappa_s$, which is defined by
\begin{equation}\label{eq:kappa_s}
\kappa_s = \frac {M_\star}{2 \pi r^2_s \Sigma_\text{crit} } ,
\end{equation}
where $M_\star$ is the total mass in stars. In order to find a realistic range for its prior, we take the stellar mass fractions estimated by \citet{2011MNRAS.417.3000S} for different initial mass functions (IMFs).
In fact, one can rewrite Eq.~\eqref{eq:kappa_s} as
\begin{equation}
\kappa_s = \frac {f_\star M_E (1+\delta)} {2 \pi r^2_s \Sigma_\text{crit} } ,
\end{equation}
where $f_\star$ and $M_E$ are the stellar mass fraction and the mass of the lens inside the Einstein radius, respectively, and $\delta$ corrects for the fact that the stellar mass profile extends well outside the Einstein radius $R_E$.
\citet{2011MNRAS.417.3000S} estimated $R_E$ = 2.5 $R_\text{eff}$, which means an estimated 28.5\% of the stellar mass is outside the Einstein radius.
Among the IMFs considered by \citet{2011MNRAS.417.3000S}, we choose those which were not shown to be in disagreement with the kinematic data: Chabrier and Salpeter. We then perform the lens analysis with a two-component lens, with a fixed Hernquist model representing the stellar component in addition to the EPL model for the dark matter distribution.
The results are shown in Table \ref{tab:hern_lens_param}.
We note that the slope~$t$ for the EPL models derived in this case are less extreme than the one we obtain for a model with a single lens component.
We also tried to get an independent constraint on the normalisation of the Hernquist component by leaving $\kappa_s$ as a free parameter, but in this case the preferred model is the one with the lowest possible $\kappa_s$ in our prior, an indication that there is no need for a two-component system to explain the lensing observables.
This is not surprising, since the region probed by the lensed images covers a small range in radius where the stellar density is very low, as can be seen from Figure \ref{fig:lens_profile}.

\begin{table*}
	\centering
	\caption{Lens parameters for the two-component model: $\kappa_s$ is the fixed normalisation of the Hernquist model for each considered IMF, the other parameters refer to the EPL dark matter distribution. All models were constructed with 5 S\'{e}rsic sources.}
	\label{tab:hern_lens_param}
	\begin{tabular}{cccccccc} 
		\hline
		IMF & $\kappa_s$ & $b$ (arcsec) & $t$ & $q$ & P.A. & $\gamma_1$ & $\gamma_2$ \\
		\hline
%	0.249 < $\kappa_s$ <  4.225 &    0.249  &    232.07  &    237.82  &      4.355  &      0.238  &      0.854  &     24.1  &      0.051  &      0.034 \\ 
 %       fixed  &  0.794  &    231.46  &    238.17  &      3.630  &      0.381  &      0.808  &     23.6  &      0.062  &      0.026 \\ 
     Chabrier & 0.9040 &      4.215 $\pm$     0.018  &        0.311 $\pm$     0.007  &        0.8380 $\pm$     0.0009  &        23.9 $\pm$      0.1  &        0.0555 $\pm$     0.0004  &        0.0303 $\pm$     0.0007 \\   
     Salpeter & 2.1954  &      3.581 $\pm$     0.025  &        0.437 $\pm$     0.009  &        0.7915 $\pm$     0.0014  &        23.3 $\pm$      0.1  &        0.0669 $\pm$     0.0006  &        0.0227 $\pm$     0.0010 \\   

		\hline
	\end{tabular}
\end{table*}

\section{Discussion} \label{sec:discussion}

\subsection{Degeneracy between lens slope and source size}\label{sec:degeneracy}
As shown in Fig.~\ref{fig:sources_caustic}, the source configurations and the caustics for the different mass models are very similar, except for a scaling of their size which depends on the slope~$t$.
This is a consequence of a transformation between models with different slopes, which is analogous to the well-known mass-sheet transformation \citep*[MST, see][]{1985ApJ...289L...1F}. The MST produces indistinguishable lensing observables if the surface density of the lens is changed as 
\begin{equation}
\kappa^\prime(\theta) = \lambda \kappa(\theta) + (1- \lambda)
\end{equation}
while the coordinates on the source plane are scaled as
\begin{equation}\label{eq:src_transf}
\beta^\prime = \lambda \beta ,
\end{equation} 
where $\theta$ and $\beta$ are the angles on the image and source plane, respectively.
Here and in the following, the lens is assumed to have circular symmetry, although the MST is valid in the general case.
Under a MST, the deflection angle $\alpha$ transforms as
\begin{equation} \label{eq:mst_alpha}
\alpha^\prime(\theta) = \lambda \alpha(\theta) + (1- \lambda)\theta
\end{equation}

The transformation between power-law models does not follow this relation, but we now show that when only a small range in the radial distance from the centre of the source is concerned, it can be approximated by a MST.
In the circularly symmetric case, the deflection angle of the EPL lens is
\begin{equation} \label{eq:epl_spherical_alpha}
\alpha(\theta) = b \left( \frac \theta b \right)^{1-t}.
\end{equation}
This can be expanded to first order around $\theta = b$ as
\begin{equation}
\alpha(\theta) \sim b \left[ 1 + (1-t) \frac {\theta - b} b \right] .
\end{equation}
For two different density slopes~$t_1$ and~$t_2$, the respective deflection angles $\alpha_1$ and $\alpha_2$ at the same angle $\theta$ from the centre are related by
\begin{equation}\label{eq:quasimst}
\alpha_2(\theta) = \frac {t_2} {t_1} \alpha_1 (\theta) + \left(1- \frac {t_2} {t_1} \right) \theta ,
\end{equation}
which is the MST in Eq.~\eqref{eq:mst_alpha} with parameter $\lambda = t_2/t_1$.
The magnifications for the two lenses are related as
\begin{equation}
\mu_2(\theta) = \frac {t_1} {t_2} \mu_1(\theta) .
\end{equation}
In our case, due to the presence of elliptical mass distributions and external shear, the relation between the different deflection angles is not exactly Eq.~\eqref{eq:quasimst}, but this relation describes in a simplified form what can be observed in our reconstructions when we change the density slope $t$.
In fact, the source morphology in the preferred EPL model closely resembles a scaled version of the morphology in the preferred SIE model, as shown in Fig.~\ref{fig:sources_caustic}.
This is in good agreement with the transformation described by Eq.~\eqref{eq:src_transf}, even though ellipticity and external shear are present in our reconstruction.
Estimating the source size from the distance between source components 1 and 5 (cf.\@ Fig.~\ref{fig:sources_caustic}), we estimate a scaling factor of $11.88$, very close to the ratio $t_\text{SIE}/t_\text{EPL} = 12.18$ between the slopes.
For the same reason, there is a difference in magnitude of $5.394$ between the two sources, which corresponds to a luminosity ratio of~${\sim} 143.8$, very close to~$12.18^2 = 148.35$. 

\begin{figure}%
\includegraphics[width=\columnwidth]{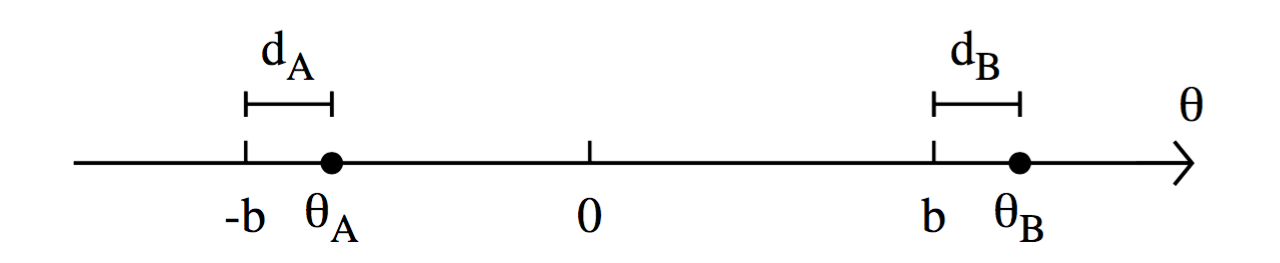}\\%
\includegraphics[width=\columnwidth]{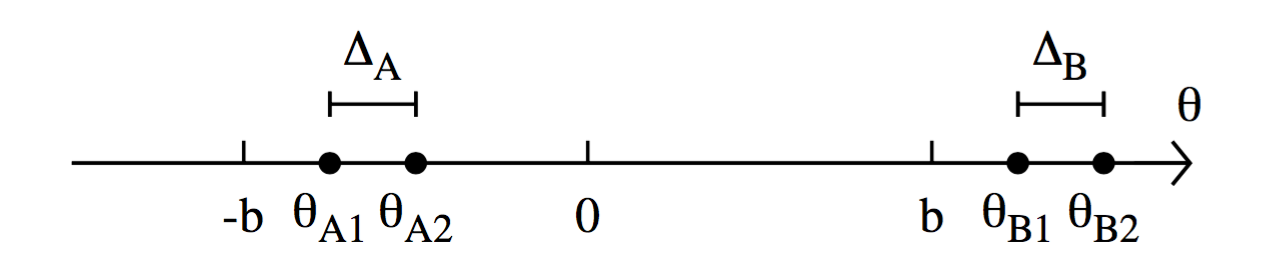}%
\caption{%
Two-image lens system from a circularly symmetric lens mass distribution.
\emph{Upper panel}: $\theta_A$ is inside the Einstein ring, $\theta_B$ is outside the Einstein ring.
The distances of the two images from the Einstein ring are $d_A$ and $d_B$, respectively.
\emph{Lower panel}: Each image is now subdivided into two components.
The distances between the components are $\Delta_A$ and $\Delta_B$, respectively.
}%
\label{fig:arrow}%
\end{figure}

Nevertheless, this degeneracy for power-law models is not complete, and this fact allows the measurement presented in Section~\ref{sect:constr_slope}.
Continuing the same argument, one can further expand Eq.~\eqref{eq:epl_spherical_alpha} to second order, for two images ($\theta_A$, $\theta_B$) of the same source located on opposite sides of the center of the lens, just inside and just outside the Einstein ring (see Fig.~\ref{fig:arrow}, top panel).
The respective deflection angles for $\theta_A$ and $\theta_B$ are, to second order,
\begin{gather}
\alpha(\theta_A) \sim b \left[ -1 + (1-t) \frac {\theta_A + b} {b}  + \frac {t(1-t)} {2} \left( \frac {\theta_A + b} {b} \right)^2 \right] , \\
\alpha(\theta_B) \sim b \left[ 1 + (1-t) \frac {\theta_B - b} {b}  - \frac {t(1-t)} {2} \left( \frac {\theta_B- b} {b} \right)^2 \right] .
\end{gather}
Defining $d_B = \theta_B - b$ and $d_A = \theta_A - ( -b) = \theta_A + b$ as the distances of the images from the Einstein ring, and identifying the source plane position for the two images, we obtain
\begin{equation}\label{eq:dist_from_re}
d_A - d_B = \frac {1-t} {2 b} (d^2_A + d^2_B) ,
\end{equation}
where it is clear that $d_A > d_B$ for $t < 1$ and $d_A <d_B$ for $t > 1$.
For a lens that is flatter than isothermal ($t < 1$), the internal image is more distant from the Einstein ring than the external one, and the opposite is true if the lens is steeper than isothermal ($t > 1$).
However, the distance from the Einstein radius in a two-image systems is not an observable, so this result cannot be used to constrain the slope of the lens.

On the other hand, if instead of a single source, we have two distinct sources (or source components) 1 and 2, the distances between the pairs of images are observable.
We name the two images inside the Einstein radius $\theta_{A1}$ and $\theta_{A2}$, and the two images outside the Einstein radius $\theta_{B!}$ and $\theta_{B2}$. The resulting configuration is the one of Figure~\ref{fig:arrow}, in the bottom panel.
We can then write the analogue of Eq.~\eqref{eq:dist_from_re} for both sources.
Subtracting one from the other, we obtain
\begin{equation}\label{eq:dist_btw_sources}
\Delta_A - \Delta_B = \frac {1-t} {2 b} (d^2_{A2}+ d^2_{B2}-d^2_{A1}-d^2_{B1}) ,
\end{equation}
where $\Delta_A = \theta_{A2}-\theta_{A1}$, $\Delta_B = \theta_{B2}-\theta_{B1}$, and the term in parentheses is by construction positive since $d_{A2} > d_{A1}$ and $d_{B2} > d_{B1}$.
Therefore, we find that the distance between the components in the image inside the Einstein ring is larger than outside for $t < 1$, and smaller for $t > 1$.
This result agrees qualitatively with Fig.~4 of \citet{2012MNRAS.426..868S}, which was obtained through numerical solution of the lens equations. 

Although the case we are examining is complicated by the presence of ellipticity and external shear, the result of our analysis has the same theoretical origin.
The constraint on the slope derives from the relative positions (cf.\@ Figs.~\ref{fig:slope_image_cont}, \ref{fig:sources_caustic}, \ref{fig:epl_images}) of source components~1 and~4 in images~B (inside the critical line) and~C (outside the critical line).
In the SIE model, the relative distances of the images B1, B4, C1 and C4 cannot be fit to the observed ones, and the best configuration the algorithm is able to find is the one where B1 is off while the other three images are in the right position.
This happens because B1 is the least luminous of these images, therefore a mismatch in the reconstruction least penalises the likelihood of the model.
For lower values of the slope~$t$, the radial distance between images~B1 and~B4 increases, and the model agrees more closely with the observation (Fig.~\ref{fig:slope_image_cont}).

\subsection{Consistency with theoretical expectations}\label{sec:slope_expectations}

\begin{figure}
 \includegraphics[width=\columnwidth]{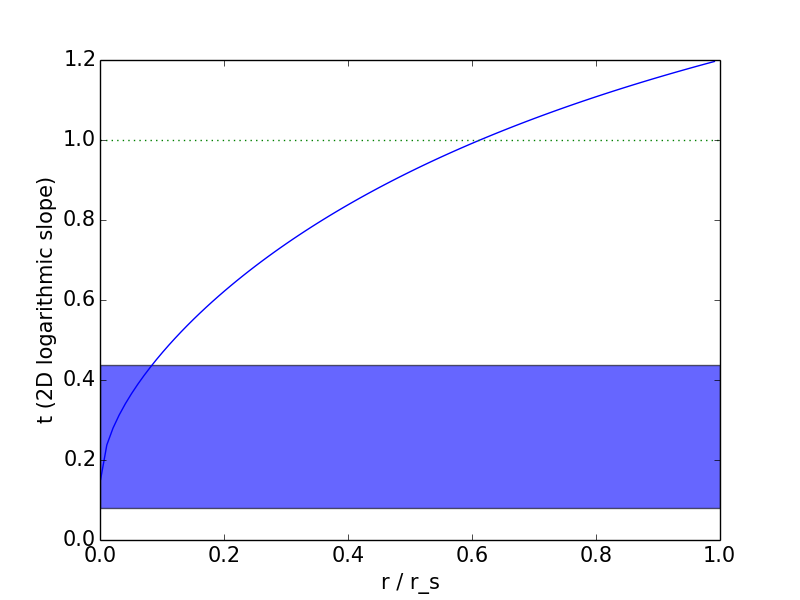}
 \caption{The projected logarithmic slope for a NFW profile as a function of radius divided be the scale radius.  The blue region marks the range of values from the lens models.  The dotted line marks the value expected for an SIS.}
 \label{fig:t_nfw}
\end{figure}

Cosmological $N$-body simulations containing only dark matter have long found that dark matter halos have a universal average profile close to NFW \citep{1997ApJ...490..493N,2004MNRAS.349.1039N}.
This profile asymptotically goes as $\rho(r) \propto r^{-1}$ for small radii.
Figure~\ref{fig:t_nfw} shows a plot of the slope of the projected surface density for this profile as a function of radius, in units of the scale radius $r_s$.
The range of values we get in the analyses of Sect.~\ref{sect:constr_slope} and~\ref{sec:two-compo} for the slope at the Einstein radius is marked in the plot.
From this we see that the data is consistent with an NFW profile if $R_E \simlt 0.09 r_s$.
Given that the physical size of $R_E$ is $\sim 29.4$~kpc, the previous constraint translates into a lower limit on the scale radius, $r_s \simgt 0.33$~Mpc for the assumed cosmology.

In order to estimate the scale radius of this object, and verify if the previous constraint is satisfied, we need an estimate of $M_{200}$. The mass estimated from lensing analysis is in this case not appropriate, because it refers only to the most internal part of the halo. Instead, we can estimate $M_{200}$ from the stellar mass of the lensing galaxy.
\citet{2011MNRAS.417.3000S} derived a stellar mass using spectroscopic data from the X-Shooter Lens Survey and dynamical modeling.
They find that the stellar mass within the Einstein radius is $10\%$ - $20\%$ of the total mass within this radius, which is $5.0\times 10^{11}~\msun$ to $M_* \sim 1.0\times 10^{12}~\msun$ depending on the kinematic model.
As already mentioned in Section~\ref{sec:two-compo}, in their model, the Einstein radius is ${\sim} 2.5$ times the effective radius of the galaxy, which for a Hernquist constant mass-to-light model means an additional $\sim 28.5\%$ of the mass lies outside the Einstein radius.
The stellar mass can be matched to the halo's $M_{200}$ by comparison with cosmological simulations.
\citet{2010ApJ...710..903M} compared the luminosity function of observed galaxies to the halo mass function of dark-matter only $N$-body simulations to find such a relation.
Using their relation, a stellar mass~$M_* \sim 1.0\times 10^{12}~\msun$ implies an unrealistically large halo mass $M_{200} \sim 10^{16}~M_\odot$, although this is beyond the mass range over which their  relation is valid.
A stellar mass~$M_* \sim 5.0\times 10^{11}~\msun$ corresponds to $M_{200} \simeq 5.3\times 10^{14}~\msun$, which is more reasonable, but still quite large.

However, more recent work with hydrodynamic cosmological simulations predicts a higher stellar mass fraction in this mass range.
The EAGLE simulations predict $M_* \sim 0.015 \, M_{200}$ \citep{2015MNRAS.451.1247S}, which gives a much more reasonable range of $M_{200} \simeq 4.5\times 10^{13}~\msun$ to $9.1\times 10^{13}~\msun$ for \citet{2011MNRAS.417.3000S}'s dynamical stellar mass estimates. The size $r_{200}$ is then in the range between 0.6 Mpc and 0.8 Mpc which implies a ratio $R_E/r_{200}$ between 0.04 and 0.05.
Using the constraint $r_s \simgt 0.33$~Mpc derived from Fig.~\ref{fig:t_nfw} gives concentrations of $c \equiv r_{200}/r_s \simeq 1.9$ to $2.4$ ($\log_{10}(c) \simeq 0.29 - 0.39$) or smaller, more concentrated for higher $M_{200}$.

The theoretical predictions for the average concentration of a halo of this mass range from $\log_{10}(c) =$ 0.69 to 0.79 \citep{2009ApJ...707..354Z,2011MNRAS.411..584M,2012MNRAS.422..185G,2016MNRAS.460.1214L} and thus the lens' halo appears to have a low concentration.  However, the predicted scatter in the concentration is significant (rms in $\log_{10}(c) \sim 0.13$  according to \citet{2012MNRAS.422..185G}) and "relaxed" halos at this mass scale with concentrations as low as $\log_{10}(c)  \sim 0.3$ are seen in the simulations (and even lower concentration non-relaxed halos) \citep{2016MNRAS.460.1214L}.
There may be some tension with theory here, but we do not consider it to be a significant inconsistency given the modeling uncertainties and the sample size of one.

The EAGLE simulations also predict that baryons will not significantly shift the slope of the mass profile at $r/r_{200} \sim 0.03$ from the NFW prediction for this halo mass scale.  The observed galaxy field surrounding the lens suggests a small group with ${\sim} 26$ smaller members \citep{2007ApJ...671L...9B}.  This is consistent with the expected halo occupation distribution (HOD) for a halo of this size \citep{2004ApJ...609...35K}.

In summary, the slope we find for the mass profile at the Einstein radius is not inconsistent with the predictions of a $\Lambda$CDM cosmology.

\begin{table*}%
\setlength{\tabcolsep}{.8\tabcolsep}%
\centering%
\caption{%
Lens parameters derived by our results and by previous published analyses.
}%
\label{tab:comp_lens_param}%
\begin{tabular}{cccccccccc} 
\hline
work                        &  lens         & source       & data               & $b$ [arcsec] & $t$    & $q$    & P.A. [\degr] & $\gamma_1$ & $\gamma_2$ \\
\hline
\citet{2008MNRAS.388..384D} & SIE+$\gamma$  & ---          & INT                & $4.98$       & $1.00$ & $0.81$ & $49.8$       & $0.016$    & $0.004$    \\
\citet{2008MNRAS.388..384D} & EPL+$\gamma$  & ---          & INT              & $5.08$       & $0.95$ & $0.83$ & $50.8$       & $0.019$    & $0.005$    \\
\citet{2011MNRAS.417.3000S} & gNFW+$\gamma$ & ---          & INT $+$ kinematics & ---          & $0.72$ & ---    & ---          & ---        & ---        \\
\citet{2013MNRAS.429L..35A} & EPL+$\gamma$  & ---          & INT $+$ kinematics & ---          & $0.76$ & ---    & ---          & ---        & ---        \\
\citet{2013ApJ...765...48J} & dPIE          & ---          & \textit{HST}                & ---          & ---    & $1.00$ & $134 $       & ---        & ---        \\
\citet{2014MNRAS.441.3238K} & SIE+$\gamma$  & 1 S\'{e}rsic & \textit{HST}               & $5.08$       & $1.00$ & $0.89$ & $40.0$       & $0.019$    & $-0.010$~~ \\
\citet{2015ApJ...814...26N} & EPL+$\gamma$  & 1 S\'{e}rsic & \textit{HST}                & $5.08$       & $0.66$ & $0.90$ & $37.1$       & $0.004$    & $-0.022$~~ \\
this work                   & SIE+$\gamma$  & 1 S\'{e}rsic & \textit{HST}                & $5.13$       & $1.00$ & $0.87$ & $40.5$       & $0.020$    & $-0.021$~~ \\
this work                   & SIE+$\gamma$  & 5 S\'{e}rsic & \textit{HST}                & $5.13$       & $1.00$ & $0.83$ & $23.4$       & $0.054$    & $-0.019$~~ \\
this work                   & EPL+$\gamma$  & 5 S\'{e}rsic & \textit{HST}                & $5.16$       & $0.08$ & $0.89$ & $24.4$       & $0.038$    & $0.038$    \\
\hline
\end{tabular}%
\end{table*}

\subsection{Comparison with previous results} 
In Table~\ref{tab:comp_lens_param} we compare our results with the ones derived by previous analyses.
We have to consider that previous results were obtained with different data, methodologies, and mass models.
Our analysis is the first one in which a detailed model was applied to the high-quality images from \textit{HST} observations.

In particular, \citet{2008MNRAS.388..384D} analysed an $i$-band observation taken with the Isaac Newton Telescope (INT) in La Palma using a semi-linear inversion technique \citep{2003ApJ...590..673W}.
This ground-based observation did not highlight all the features of the arcs visible in the \textit{HST} image.
However, the source and image reconstructions they show resemble a low-resolution version of our Figs.~\ref{fig:sources_caustic} and~\ref{fig:epl_images}, which contain only two source components, the one we label 1 and the other a combination of 2, 3 and 4.
In contrast with our results, \citet{2008MNRAS.388..384D} find a very nearly isothermal slope for the lens ($t = 0.95$).

Subsequently, \citet{2011MNRAS.417.3000S} and \citet{2013MNRAS.429L..35A} used the constraint from \citet{2008MNRAS.388..384D} on the mass within the Einstein radius, and added stellar kinematics of the host galaxy to estimate the mass distribution at smaller radii.
Even if \citet{2011MNRAS.417.3000S} formally used a generalised NFW model, they fixed the scale radius to 50~kpc, which is ${\sim} 2 R_E$, so we can compare the internal slope they get to our value for $t$.
Both works agree with this one in getting a flatter-than-isothermal slope, even if the value is significantly different.
We underline that their analysis included kinematics data, while our results is derived from lensing only. 

The analysis by \citet{2013ApJ...765...48J} is one of two works that, to our knowledge, analyse \textit{HST} data.
They used the Lenstool program \citep{2007NJPh....9..447J}, which -- in contrast to our full-image approach -- takes only the positions of multiply imaged background sources as constraints.
The lens model they employed is a double pseudo-isothermal elliptical mass distribution \citep{1993ApJ...417..450K}, the slope of which changes at two radii $r_{core}$ and $r_{cut}$, thus making it hard to compare their results to ours.
Nevertheless, the fact that they get a perfect circularly symmetric lens is in contrast with our results, as well as with other analyses.
This may be due to the different technique for lens reconstruction, which does not take all the information present in the image into account.

\citet{2014MNRAS.441.3238K} used a methodology similar to ours and, perhaps unsurprisingly, their results are in good agreement with ours, if we consider the same model for source and lens.
This happens even though they used a lower-resolution image of the object, taken from SDSS data.
This supports our idea that the most simple version of our model (SIE+$\gamma$, 1 S\'{e}rsic) would be enough to model available ground-based data, where the distinctive features of the different multiple images are not resolved.
The reconstruction of space-based images requires a finer treatment of the source, and while this is demanding in terms of computing time and efficiency of the algorithm, it gives a greater insight into the structure of the lens, as we showed in the previous sections.

Finally, \citet{2015ApJ...814...26N} applied a parametric modelling technique to \textit{HST} multi-band data.
We quote in Table~\ref{tab:comp_lens_param} their results from lensing only, although their work includes an analysis of kinematic data as well, which supports their estimate of the lens slope.
Their results appear to be in partial agreement with ours, as they get a flatter-than-isothermal slope.
We note, however, that their constraint on the slope is derived with a single S\'{e}rsic source component, so its origin is likely to be fundamentally different from the one we presented in Section~\ref{sect:constr_slope}, which depends on the multi-component nature of the source model.

\subsection{Group satellites and substructures}

In the analysis we presented, we did not explicitly consider the presence of structures other than the main halo, and we modelled all possible contributions of objects along the line of sight only as an external shear. Since the discovery, the Cosmic Horseshoe has been described as embedded in a group of galaxies, which is coherent with the large image separation (${\sim}~10^{\prime\prime}$) and with the mass estimate ($4.5\times 10^{13}~\msun$ to $9.1\times 10^{13}~\msun$) we derived in Section \ref{sec:slope_expectations}. At the same time, it has been noted that the image structure shows no sign of distortions due to other galaxies, because the shape is exceptionally circular \citep{2007ApJ...671L...9B,2008MNRAS.388..384D}. For this reason, all the works in the literature do not model individually any other object apart from the main galaxy.
The only exception is \citet{2013ApJ...765...48J}, although it does not state whether the inclusion of additional galaxies has any effect on the main lens parameters. The quality of the reconstruction which we obtained with a single halo, shown in Figure \ref{fig:image_and_models}, is an additional indication that the contribution from other halos is  negligible at the level of our reconstruction.

Nevertheless, the small mismatch between images shown in Section \ref{sect:constr_slope} could in principle be solved by additional components on the lens plane, instead of a different lens slope. We investigated this possibility, but unfortunately there is no visible sign of the presence of a substructure close enough to the B1 image. We tried anyway to reduce the mismatch between the SIE reconstruction and the image by adding new lens components in the locations of the few visible objects close to the main galaxy, but we did not get any improvement. There remains the possibility of trying to directly identify dark objects from the image configuration following the approach of \citet{2010MNRAS.408.1969V}, but this is outside the scope of this paper.

\section{Conclusions} \label{sec:conclusion}

The degeneracy between source size and density slope of the mass distribution is well-known, and it is common to use other observables such as the velocity dispersion of the galaxy stars to measure the mass enclosed in different radii, and thus build a mass profile \citep[see e.g.][]{2002MNRAS.337L...6T, 2007ApJ...666..726B,2011MNRAS.417.3000S,2013ApJ...777...98S}.
However, it has been pointed out that it is possible to break this degeneracy from lensing observations only, making use of extended images \citep{2008MNRAS.388..384D, 2010ApJ...711..201S, 2012MNRAS.426..868S, 2012MNRAS.427.1918E}.   

The Horseshoe lens is exceptional in that the complexity of the source is well resolved in the available data.
This has allowed us to attempt this measurement with our flexible and efficient lens reconstruction code \textsc{Lensed}.
The results are two-fold.
We do get a constraint on the density slope, but we also noticed that in a real-world situation, where source and lens are not expected to precisely follow any simple assumptions, overcoming the degeneracy is very challenging.
Great caution must be taken to understand if the constraint is really due to the data or perhaps comes from some systematics between the model and the algorithm. 

We investigated the origin of our constraint and revealed the image details that drove the algorithm towards a flatter-than-isothermal slope. 
We studied with an analytical approximation both the aforementioned degeneracy and how it is broken by using the relative distance between sub-components of the source in the different arcs.
We also verified that the slope values we get at the Einstein radius are compatible with a NFW profile, if $R_{E} \sim 0.1 r_s$, which is realistic given the mass estimate of the object.

We finally note that this uncertainty on the slope of the mass distribution may pose fundamental problems to many studies which use galaxy strong lenses as lensing telescopes to investigate the properties of high-redshift sources.
On the other hand, if one one can put a motivated prior on the source size from external arguments, this reduces the uncertainty on the lens, as shown by \citet{2015arXiv151103662B}.

\section*{Acknowledgements}

This research is part of project GLENCO, funded under the European Seventh Framework Programme, Ideas, Grant Agreement n.~259349.

The analysis presented in this paper is based on observations made with the NASA/ESA \textit{Hubble Space Telescope}, and obtained from the Hubble Legacy Archive, which is a collaboration between the Space Telescope Science Institute (STScI/NASA), the Space Telescope European Coordinating Facility (ST-ECF/ESA) and the Canadian Astronomy Data Centre (CADC/NRC/CSA).

\bibliographystyle{mnras}
\bibliography{horseshoe}

\begin{thebibliography}{}
\makeatletter
\relax
\def\mn@urlcharsother{\let\do\@makeother \do\$\do\&\do\#\do\^\do\_\do\%\do\~}
\def\mn@doi{\begingroup\mn@urlcharsother \@ifnextchar [ {\mn@doi@}
  {\mn@doi@[]}}
\def\mn@doi@[#1]#2{\def\@tempa{#1}\ifx\@tempa\@empty \href
  {http://dx.doi.org/#2} {doi:#2}\else \href {http://dx.doi.org/#2} {#1}\fi
  \endgroup}
\def\mn@eprint#1#2{\mn@eprint@#1:#2::\@nil}
\def\mn@eprint@arXiv#1{\href {http://arxiv.org/abs/#1} {{\tt arXiv:#1}}}
\def\mn@eprint@dblp#1{\href {http://dblp.uni-trier.de/rec/bibtex/#1.xml}
  {dblp:#1}}
\def\mn@eprint@#1:#2:#3:#4\@nil{\def\@tempa {#1}\def\@tempb {#2}\def\@tempc
  {#3}\ifx \@tempc \@empty \let \@tempc \@tempb \let \@tempb \@tempa \fi \ifx
  \@tempb \@empty \def\@tempb {arXiv}\fi \@ifundefined
  {mn@eprint@\@tempb}{\@tempb:\@tempc}{\expandafter \expandafter \csname
  mn@eprint@\@tempb\endcsname \expandafter{\@tempc}}}

\bibitem[\protect\citeauthoryear{{Agnello}, {Auger}  \& {Evans}}{{Agnello}
  et~al.}{2013}]{2013MNRAS.429L..35A}
{Agnello} A.,  {Auger} M.~W.,   {Evans} N.~W.,  2013, \mn@doi [\mnras]
  {10.1093/mnrasl/sls020}, \href
  {http://adsabs.harvard.edu/abs/2013MNRAS.429L..35A} {429, L35}

\bibitem[\protect\citeauthoryear{{Auger}, {Treu}, {Bolton}, {Gavazzi},
  {Koopmans}, {Marshall}, {Moustakas}  \& {Burles}}{{Auger}
  et~al.}{2010}]{2010ApJ...724..511A}
{Auger} M.~W.,  {Treu} T.,  {Bolton} A.~S.,  {Gavazzi} R.,  {Koopmans}
  L.~V.~E.,  {Marshall} P.~J.,  {Moustakas} L.~A.,   {Burles} S.,  2010,
  \mn@doi [\apj] {10.1088/0004-637X/724/1/511}, \href
  {http://adsabs.harvard.edu/abs/2010ApJ...724..511A} {724, 511}

\bibitem[\protect\citeauthoryear{{Barnab{\`e}} \& {Koopmans}}{{Barnab{\`e}} \&
  {Koopmans}}{2007}]{2007ApJ...666..726B}
{Barnab{\`e}} M.,  {Koopmans} L.~V.~E.,  2007, \mn@doi [\apj] {10.1086/520495},
  \href {http://adsabs.harvard.edu/abs/2007ApJ...666..726B} {666, 726}

\bibitem[\protect\citeauthoryear{{Belokurov} et~al.,}{{Belokurov}
  et~al.}{2007}]{2007ApJ...671L...9B}
{Belokurov} V.,  et~al., 2007, \mn@doi [\apjl] {10.1086/524948}, \href
  {http://adsabs.harvard.edu/abs/2007ApJ...671L...9B} {671, L9}

\bibitem[\protect\citeauthoryear{{Birrer}, {Amara}  \& {Refregier}}{{Birrer}
  et~al.}{2015}]{2015arXiv151103662B}
{Birrer} S.,  {Amara} A.,   {Refregier} A.,  2015, preprint, \href
  {http://adsabs.harvard.edu/abs/2015arXiv151103662B} {} (\mn@eprint {arXiv}
  {1511.03662})

\bibitem[\protect\citeauthoryear{{Bolton}, {Burles}, {Koopmans}, {Treu}  \&
  {Moustakas}}{{Bolton} et~al.}{2006}]{2006ApJ...638..703B}
{Bolton} A.~S.,  {Burles} S.,  {Koopmans} L.~V.~E.,  {Treu} T.,   {Moustakas}
  L.~A.,  2006, \mn@doi [\apj] {10.1086/498884}, \href
  {http://adsabs.harvard.edu/abs/2006ApJ...638..703B} {638, 703}

\bibitem[\protect\citeauthoryear{Burkert}{Burkert}{1995}]{1538-4357-447-1-L25}
Burkert A.,  1995, \apjl, 447, L25

\bibitem[\protect\citeauthoryear{{Cappellari} et~al.,}{{Cappellari}
  et~al.}{2015}]{2015ApJ...804L..21C}
{Cappellari} M.,  et~al., 2015, \mn@doi [\apjl] {10.1088/2041-8205/804/1/L21},
  \href {http://esoads.eso.org/abs/2015ApJ...804L..21C} {804, L21}

\bibitem[\protect\citeauthoryear{{Chae}, {Bernardi}  \& {Kravtsov}}{{Chae}
  et~al.}{2014}]{2014MNRAS.437.3670C}
{Chae} K.-H.,  {Bernardi} M.,   {Kravtsov} A.~V.,  2014, \mn@doi [\mnras]
  {10.1093/mnras/stt2163}, \href
  {http://esoads.eso.org/abs/2014MNRAS.437.3670C} {437, 3670}

\bibitem[\protect\citeauthoryear{Dalcanton \& Hogan}{Dalcanton \&
  Hogan}{2001}]{0004-637X-561-1-35}
Dalcanton J.~J.,  Hogan C.~J.,  2001, \apj, 561, 35

\bibitem[\protect\citeauthoryear{{Dye}, {Evans}, {Belokurov}, {Warren}  \&
  {Hewett}}{{Dye} et~al.}{2008}]{2008MNRAS.388..384D}
{Dye} S.,  {Evans} N.~W.,  {Belokurov} V.,  {Warren} S.~J.,   {Hewett} P.,
  2008, \mn@doi [\mnras] {10.1111/j.1365-2966.2008.13401.x}, \href
  {http://adsabs.harvard.edu/abs/2008MNRAS.388..384D} {388, 384}

\bibitem[\protect\citeauthoryear{{Eichner}, {Seitz}  \& {Bauer}}{{Eichner}
  et~al.}{2012}]{2012MNRAS.427.1918E}
{Eichner} T.,  {Seitz} S.,   {Bauer} A.,  2012, \mn@doi [\mnras]
  {10.1111/j.1365-2966.2012.22003.x}, \href
  {http://adsabs.harvard.edu/abs/2012MNRAS.427.1918E} {427, 1918}

\bibitem[\protect\citeauthoryear{{Falco}, {Gorenstein}  \& {Shapiro}}{{Falco}
  et~al.}{1985}]{1985ApJ...289L...1F}
{Falco} E.~E.,  {Gorenstein} M.~V.,   {Shapiro} I.~I.,  1985, \mn@doi [\apjl]
  {10.1086/184422}, \href {http://adsabs.harvard.edu/abs/1985ApJ...289L...1F}
  {289, L1}

\bibitem[\protect\citeauthoryear{{Feroz} \& {Hobson}}{{Feroz} \&
  {Hobson}}{2008}]{2008MNRAS.384..449F}
{Feroz} F.,  {Hobson} M.~P.,  2008, \mn@doi [\mnras]
  {10.1111/j.1365-2966.2007.12353.x}, \href
  {http://adsabs.harvard.edu/abs/2008MNRAS.384..449F} {384, 449}

\bibitem[\protect\citeauthoryear{{Feroz}, {Hobson}  \& {Bridges}}{{Feroz}
  et~al.}{2009}]{2009MNRAS.398.1601F}
{Feroz} F.,  {Hobson} M.~P.,   {Bridges} M.,  2009, \mn@doi [\mnras]
  {10.1111/j.1365-2966.2009.14548.x}, \href
  {http://adsabs.harvard.edu/abs/2009MNRAS.398.1601F} {398, 1601}

\bibitem[\protect\citeauthoryear{{Feroz}, {Hobson}, {Cameron}  \&
  {Pettitt}}{{Feroz} et~al.}{2013}]{2013arXiv1306.2144F}
{Feroz} F.,  {Hobson} M.~P.,  {Cameron} E.,   {Pettitt} A.~N.,  2013, preprint,
  \href {http://adsabs.harvard.edu/abs/2013arXiv1306.2144F} {} (\mn@eprint
  {arXiv} {1306.2144})

\bibitem[\protect\citeauthoryear{{Gavazzi}, {Treu}, {Koopmans}, {Bolton},
  {Moustakas}, {Burles}  \& {Marshall}}{{Gavazzi}
  et~al.}{2008}]{2008ApJ...677.1046G}
{Gavazzi} R.,  {Treu} T.,  {Koopmans} L.~V.~E.,  {Bolton} A.~S.,  {Moustakas}
  L.~A.,  {Burles} S.,   {Marshall} P.~J.,  2008, \mn@doi [\apj]
  {10.1086/529541}, \href {http://adsabs.harvard.edu/abs/2008ApJ...677.1046G}
  {677, 1046}

\bibitem[\protect\citeauthoryear{{Gilmore}, {Wilkinson}, {Wyse}, {Kleyna},
  {Koch}, {Evans}  \& {Grebel}}{{Gilmore} et~al.}{2007}]{2007ApJ...663..948G}
{Gilmore} G.,  {Wilkinson} M.~I.,  {Wyse} R.~F.~G.,  {Kleyna} J.~T.,  {Koch}
  A.,  {Evans} N.~W.,   {Grebel} E.~K.,  2007, \mn@doi [\apj] {10.1086/518025},
  \href {http://esoads.eso.org/abs/2007ApJ...663..948G} {663, 948}

\bibitem[\protect\citeauthoryear{{Giocoli}, {Tormen}  \& {Sheth}}{{Giocoli}
  et~al.}{2012}]{2012MNRAS.422..185G}
{Giocoli} C.,  {Tormen} G.,   {Sheth} R.~K.,  2012, \mn@doi [\mnras]
  {10.1111/j.1365-2966.2012.20594.x}, \href
  {http://adsabs.harvard.edu/abs/2012MNRAS.422..185G} {422, 185}

\bibitem[\protect\citeauthoryear{{Hernquist}}{{Hernquist}}{1990}]{1990ApJ...356..359H}
{Hernquist} L.,  1990, \mn@doi [\apj] {10.1086/168845}, \href
  {http://adsabs.harvard.edu/abs/1990ApJ...356..359H} {356, 359}

\bibitem[\protect\citeauthoryear{Jing \& Suto}{Jing \&
  Suto}{2000}]{1538-4357-529-2-L69}
Jing Y.~P.,  Suto Y.,  2000, \apjl, 529, L69

\bibitem[\protect\citeauthoryear{{Jones}, {Ellis}, {Richard}  \&
  {Jullo}}{{Jones} et~al.}{2013}]{2013ApJ...765...48J}
{Jones} T.,  {Ellis} R.~S.,  {Richard} J.,   {Jullo} E.,  2013, \mn@doi [\apj]
  {10.1088/0004-637X/765/1/48}, \href
  {http://adsabs.harvard.edu/abs/2013ApJ...765...48J} {765, 48}

\bibitem[\protect\citeauthoryear{{Jullo}, {Kneib}, {Limousin},
  {El{\'{\i}}asd{\'o}ttir}, {Marshall}  \& {Verdugo}}{{Jullo}
  et~al.}{2007}]{2007NJPh....9..447J}
{Jullo} E.,  {Kneib} J.-P.,  {Limousin} M.,  {El{\'{\i}}asd{\'o}ttir} {\'A}.,
  {Marshall} P.~J.,   {Verdugo} T.,  2007, \mn@doi [New Journal of Physics]
  {10.1088/1367-2630/9/12/447}, \href
  {http://adsabs.harvard.edu/abs/2007NJPh....9..447J} {9, 447}

\bibitem[\protect\citeauthoryear{{Kassiola} \& {Kovner}}{{Kassiola} \&
  {Kovner}}{1993}]{1993ApJ...417..450K}
{Kassiola} A.,  {Kovner} I.,  1993, \mn@doi [\apj] {10.1086/173325}, \href
  {http://adsabs.harvard.edu/abs/1993ApJ...417..450K} {417, 450}

\bibitem[\protect\citeauthoryear{{Kochanek}, {Schneider}  \&
  {Wambsganss}}{{Kochanek} et~al.}{2004}]{2006glsw.conf.....M}
{Kochanek} C.~S.,  {Schneider} P.,   {Wambsganss} J.,  2004, Part 2 of
  Gravitational Lensing: Strong, Weak \& Micro, Proceedings of the 33rd
  Saas-Fee Advanced Course.
G. Meylan, P. Jetzer \& P. North, eds. (Springer-Verlag: Berlin)

\bibitem[\protect\citeauthoryear{{Kormann}, {Schneider}  \&
  {Bartelmann}}{{Kormann} et~al.}{1994}]{1994A&A...284..285K}
{Kormann} R.,  {Schneider} P.,   {Bartelmann} M.,  1994, \aap, \href
  {http://adsabs.harvard.edu/abs/1994A%26A...284..285K} {284, 285}

\bibitem[\protect\citeauthoryear{{Kostrzewa-Rutkowska}, {Wyrzykowski}, {Auger},
  {Collett}  \& {Belokurov}}{{Kostrzewa-Rutkowska}
  et~al.}{2014}]{2014MNRAS.441.3238K}
{Kostrzewa-Rutkowska} Z.,  {Wyrzykowski} {\L}.,  {Auger} M.~W.,  {Collett}
  T.~E.,   {Belokurov} V.,  2014, \mn@doi [\mnras] {10.1093/mnras/stu783},
  \href {http://adsabs.harvard.edu/abs/2014MNRAS.441.3238K} {441, 3238}

\bibitem[\protect\citeauthoryear{{Kravtsov}, {Berlind}, {Wechsler}, {Klypin},
  {Gottl{\"o}ber}, {Allgood}  \& {Primack}}{{Kravtsov}
  et~al.}{2004}]{2004ApJ...609...35K}
{Kravtsov} A.~V.,  {Berlind} A.~A.,  {Wechsler} R.~H.,  {Klypin} A.~A.,
  {Gottl{\"o}ber} S.,  {Allgood} B.,   {Primack} J.~R.,  2004, \mn@doi [\apj]
  {10.1086/420959}, \href {http://adsabs.harvard.edu/abs/2004ApJ...609...35K}
  {609, 35}

\bibitem[\protect\citeauthoryear{{Krist}, {Hook}  \& {Stoehr}}{{Krist}
  et~al.}{2011}]{2011SPIE.8127E..0JK}
{Krist} J.~E.,  {Hook} R.~N.,   {Stoehr} F.,  2011, in Optical Modeling and
  Performance Predictions V. p. 81270J, \mn@doi{10.1117/12.892762}

\bibitem[\protect\citeauthoryear{Laporte \& White}{Laporte \&
  White}{2015}]{Laporte01082015}
Laporte C. F.~P.,  White S. D.~M.,  2015, \mn@doi [\mnras]
  {10.1093/mnras/stv112}, 451, 1177

\bibitem[\protect\citeauthoryear{{Leier}, {Ferreras}, {Saha}, {Charlot},
  {Bruzual}  \& {La Barbera}}{{Leier} et~al.}{2016}]{2016MNRAS.tmp..677L}
{Leier} D.,  {Ferreras} I.,  {Saha} P.,  {Charlot} S.,  {Bruzual} G.,   {La
  Barbera} F.,  2016, \mn@doi [\mnras] {10.1093/mnras/stw885}, \href
  {http://adsabs.harvard.edu/abs/2016MNRAS.tmp..677L} {}

\bibitem[\protect\citeauthoryear{{Ludlow}, {Bose}, {Angulo}, {Wang},
  {Hellwing}, {Navarro}, {Cole}  \& {Frenk}}{{Ludlow}
  et~al.}{2016}]{2016MNRAS.460.1214L}
{Ludlow} A.~D.,  {Bose} S.,  {Angulo} R.~E.,  {Wang} L.,  {Hellwing} W.~A.,
  {Navarro} J.~F.,  {Cole} S.,   {Frenk} C.~S.,  2016, \mn@doi [\mnras]
  {10.1093/mnras/stw1046}, \href
  {http://adsabs.harvard.edu/abs/2016MNRAS.460.1214L} {460, 1214}

\bibitem[\protect\citeauthoryear{{Mandelbaum}, {Seljak}  \&
  {Hirata}}{{Mandelbaum} et~al.}{2008}]{2008JCAP...08..006M}
{Mandelbaum} R.,  {Seljak} U.,   {Hirata} C.~M.,  2008, \mn@doi [\jcap]
  {10.1088/1475-7516/2008/08/006}, \href
  {http://adsabs.harvard.edu/abs/2008JCAP...08..006M} {8, 006}

\bibitem[\protect\citeauthoryear{{Marshall} et~al.,}{{Marshall}
  et~al.}{2007}]{2007ApJ...671.1196M}
{Marshall} P.~J.,  et~al., 2007, \mn@doi [\apj] {10.1086/523091}, \href
  {http://adsabs.harvard.edu/abs/2007ApJ...671.1196M} {671, 1196}

\bibitem[\protect\citeauthoryear{{Moster}, {Somerville}, {Maulbetsch}, {van den
  Bosch}, {Macci{\`o}}, {Naab}  \& {Oser}}{{Moster}
  et~al.}{2010}]{2010ApJ...710..903M}
{Moster} B.~P.,  {Somerville} R.~S.,  {Maulbetsch} C.,  {van den Bosch} F.~C.,
  {Macci{\`o}} A.~V.,  {Naab} T.,   {Oser} L.,  2010, \mn@doi [\apj]
  {10.1088/0004-637X/710/2/903}, \href
  {http://adsabs.harvard.edu/abs/2010ApJ...710..903M} {710, 903}

\bibitem[\protect\citeauthoryear{{Mu{\~n}oz-Cuartas}, {Macci{\`o}},
  {Gottl{\"o}ber}  \& {Dutton}}{{Mu{\~n}oz-Cuartas}
  et~al.}{2011}]{2011MNRAS.411..584M}
{Mu{\~n}oz-Cuartas} J.~C.,  {Macci{\`o}} A.~V.,  {Gottl{\"o}ber} S.,   {Dutton}
  A.~A.,  2011, \mn@doi [\mnras] {10.1111/j.1365-2966.2010.17704.x}, \href
  {http://adsabs.harvard.edu/abs/2011MNRAS.411..584M} {411, 584}

\bibitem[\protect\citeauthoryear{{Navarro}, {Frenk}  \& {White}}{{Navarro}
  et~al.}{1997}]{1997ApJ...490..493N}
{Navarro} J.~F.,  {Frenk} C.~S.,   {White} S.~D.~M.,  1997, \apj, \href
  {http://adsabs.harvard.edu/abs/1997ApJ...490..493N} {490, 493}

\bibitem[\protect\citeauthoryear{{Navarro} et~al.,}{{Navarro}
  et~al.}{2004}]{2004MNRAS.349.1039N}
{Navarro} J.~F.,  et~al., 2004, \mn@doi [\mnras]
  {10.1111/j.1365-2966.2004.07586.x}, \href
  {http://adsabs.harvard.edu/abs/2004MNRAS.349.1039N} {349, 1039}

\bibitem[\protect\citeauthoryear{{Newman}, {Treu}, {Ellis}  \& {Sand}}{{Newman}
  et~al.}{2013}]{2013ApJ...765...25N}
{Newman} A.~B.,  {Treu} T.,  {Ellis} R.~S.,   {Sand} D.~J.,  2013, \mn@doi
  [\apj] {10.1088/0004-637X/765/1/25}, \href
  {http://esoads.eso.org/abs/2013ApJ...765...25N} {765, 25}

\bibitem[\protect\citeauthoryear{{Newman}, {Ellis}  \& {Treu}}{{Newman}
  et~al.}{2015}]{2015ApJ...814...26N}
{Newman} A.~B.,  {Ellis} R.~S.,   {Treu} T.,  2015, \mn@doi [\apj]
  {10.1088/0004-637X/814/1/26}, \href
  {http://adsabs.harvard.edu/abs/2015ApJ...814...26N} {814, 26}

\bibitem[\protect\citeauthoryear{Oguri, Rusu  \& Falco}{Oguri
  et~al.}{2014}]{Oguri11042014}
Oguri M.,  Rusu C.~E.,   Falco E.~E.,  2014, \mn@doi [\mnras]
  {10.1093/mnras/stu106}, 439, 2494

\bibitem[\protect\citeauthoryear{{Pontzen} \& {Governato}}{{Pontzen} \&
  {Governato}}{2014}]{2014Natur.506..171P}
{Pontzen} A.,  {Governato} F.,  2014, \mn@doi [\nat] {10.1038/nature12953},
  \href {http://adsabs.harvard.edu/abs/2014Natur.506..171P} {506, 171}

\bibitem[\protect\citeauthoryear{{Quider}, {Pettini}, {Shapley}  \&
  {Steidel}}{{Quider} et~al.}{2009}]{2009MNRAS.398.1263Q}
{Quider} A.~M.,  {Pettini} M.,  {Shapley} A.~E.,   {Steidel} C.~C.,  2009,
  \mn@doi [\mnras] {10.1111/j.1365-2966.2009.15234.x}, \href
  {http://adsabs.harvard.edu/abs/2009MNRAS.398.1263Q} {398, 1263}

\bibitem[\protect\citeauthoryear{{Sand}, {Treu}  \& {Ellis}}{{Sand}
  et~al.}{2002}]{2002ApJ...574L.129S}
{Sand} D.~J.,  {Treu} T.,   {Ellis} R.~S.,  2002, \mn@doi [\apjl]
  {10.1086/342530}, \href {http://esoads.eso.org/abs/2002ApJ...574L.129S} {574,
  L129}

\bibitem[\protect\citeauthoryear{{Schaller} et~al.,}{{Schaller}
  et~al.}{2015}]{2015MNRAS.451.1247S}
{Schaller} M.,  et~al., 2015, \mn@doi [\mnras] {10.1093/mnras/stv1067}, \href
  {http://adsabs.harvard.edu/abs/2015MNRAS.451.1247S} {451, 1247}

\bibitem[\protect\citeauthoryear{{Schneider} \& {Sluse}}{{Schneider} \&
  {Sluse}}{2014}]{2014A&A...564A.103S}
{Schneider} P.,  {Sluse} D.,  2014, \mn@doi [\aap]
  {10.1051/0004-6361/201322106}, \href
  {http://adsabs.harvard.edu/abs/2014A%26A...564A.103S} {564, A103}

\bibitem[\protect\citeauthoryear{{Sonnenfeld}, {Treu}, {Gavazzi}, {Marshall},
  {Auger}, {Suyu}, {Koopmans}  \& {Bolton}}{{Sonnenfeld}
  et~al.}{2012}]{2012ApJ...752..163S}
{Sonnenfeld} A.,  {Treu} T.,  {Gavazzi} R.,  {Marshall} P.~J.,  {Auger} M.~W.,
  {Suyu} S.~H.,  {Koopmans} L.~V.~E.,   {Bolton} A.~S.,  2012, \mn@doi [\apj]
  {10.1088/0004-637X/752/2/163}, \href
  {http://adsabs.harvard.edu/abs/2012ApJ...752..163S} {752, 163}

\bibitem[\protect\citeauthoryear{{Sonnenfeld}, {Treu}, {Gavazzi}, {Suyu},
  {Marshall}, {Auger}  \& {Nipoti}}{{Sonnenfeld}
  et~al.}{2013}]{2013ApJ...777...98S}
{Sonnenfeld} A.,  {Treu} T.,  {Gavazzi} R.,  {Suyu} S.~H.,  {Marshall} P.~J.,
  {Auger} M.~W.,   {Nipoti} C.,  2013, \mn@doi [\apj]
  {10.1088/0004-637X/777/2/98}, \href
  {http://adsabs.harvard.edu/abs/2013ApJ...777...98S} {777, 98}

\bibitem[\protect\citeauthoryear{{Spiniello}, {Koopmans}, {Trager}, {Czoske}
  \& {Treu}}{{Spiniello} et~al.}{2011}]{2011MNRAS.417.3000S}
{Spiniello} C.,  {Koopmans} L.~V.~E.,  {Trager} S.~C.,  {Czoske} O.,   {Treu}
  T.,  2011, \mn@doi [\mnras] {10.1111/j.1365-2966.2011.19458.x}, \href
  {http://adsabs.harvard.edu/abs/2011MNRAS.417.3000S} {417, 3000}

\bibitem[\protect\citeauthoryear{{Suyu}}{{Suyu}}{2012}]{2012MNRAS.426..868S}
{Suyu} S.~H.,  2012, \mn@doi [\mnras] {10.1111/j.1365-2966.2012.21661.x}, \href
  {http://adsabs.harvard.edu/abs/2012MNRAS.426..868S} {426, 868}

\bibitem[\protect\citeauthoryear{{Suyu}, {Marshall}, {Auger}, {Hilbert},
  {Blandford}, {Koopmans}, {Fassnacht}  \& {Treu}}{{Suyu}
  et~al.}{2010}]{2010ApJ...711..201S}
{Suyu} S.~H.,  {Marshall} P.~J.,  {Auger} M.~W.,  {Hilbert} S.,  {Blandford}
  R.~D.,  {Koopmans} L.~V.~E.,  {Fassnacht} C.~D.,   {Treu} T.,  2010, \mn@doi
  [\apj] {10.1088/0004-637X/711/1/201}, \href
  {http://adsabs.harvard.edu/abs/2010ApJ...711..201S} {711, 201}

\bibitem[\protect\citeauthoryear{{Tessore} \& {Metcalf}}{{Tessore} \&
  {Metcalf}}{2015}]{2015A&A...580A..79T}
{Tessore} N.,  {Metcalf} R.~B.,  2015, \mn@doi [\aap]
  {10.1051/0004-6361/201526773}, \href
  {http://adsabs.harvard.edu/abs/2015A%26A...580A..79T} {580, A79}

\bibitem[\protect\citeauthoryear{{Tessore}, {Bellagamba}  \&
  {Metcalf}}{{Tessore} et~al.}{2015}]{2015arXiv150507674T}
{Tessore} N.,  {Bellagamba} F.,   {Metcalf} R.~B.,  2015, preprint, \href
  {http://adsabs.harvard.edu/abs/2015arXiv150507674T} {} (\mn@eprint {arXiv}
  {1505.07674})

\bibitem[\protect\citeauthoryear{{Treu} \& {Koopmans}}{{Treu} \&
  {Koopmans}}{2002}]{2002MNRAS.337L...6T}
{Treu} T.,  {Koopmans} L.~V.~E.,  2002, \mn@doi [\mnras]
  {10.1046/j.1365-8711.2002.06107.x}, \href
  {http://adsabs.harvard.edu/abs/2002MNRAS.337L...6T} {337, L6}

\bibitem[\protect\citeauthoryear{{Vegetti}, {Koopmans}, {Bolton}, {Treu}  \&
  {Gavazzi}}{{Vegetti} et~al.}{2010}]{2010MNRAS.408.1969V}
{Vegetti} S.,  {Koopmans} L.~V.~E.,  {Bolton} A.,  {Treu} T.,   {Gavazzi} R.,
  2010, \mn@doi [\mnras] {10.1111/j.1365-2966.2010.16865.x}, \href
  {http://adsabs.harvard.edu/abs/2010MNRAS.408.1969V} {408, 1969}

\bibitem[\protect\citeauthoryear{{Warren} \& {Dye}}{{Warren} \&
  {Dye}}{2003}]{2003ApJ...590..673W}
{Warren} S.~J.,  {Dye} S.,  2003, \mn@doi [\apj] {10.1086/375132}, \href
  {http://adsabs.harvard.edu/abs/2003ApJ...590..673W} {590, 673}

\bibitem[\protect\citeauthoryear{Weinberg, Bullock, Governato, Kuzio~de Naray
  \& Peter}{Weinberg et~al.}{2015}]{Weinberg06102015}
Weinberg D.~H.,  Bullock J.~S.,  Governato F.,  Kuzio~de Naray R.,   Peter A.
  H.~G.,  2015, \mn@doi [PNAS] {10.1073/pnas.1308716112}, 112, 12249

\bibitem[\protect\citeauthoryear{{Zhao}, {Jing}, {Mo}  \& {B{\"o}rner}}{{Zhao}
  et~al.}{2009}]{2009ApJ...707..354Z}
{Zhao} D.~H.,  {Jing} Y.~P.,  {Mo} H.~J.,   {B{\"o}rner} G.,  2009, \mn@doi
  [\apj] {10.1088/0004-637X/707/1/354}, \href
  {http://adsabs.harvard.edu/abs/2009ApJ...707..354Z} {707, 354}

\bibitem[\protect\citeauthoryear{{de Vaucouleurs}}{{de
  Vaucouleurs}}{1948}]{1948AnAp...11..247D}
{de Vaucouleurs} G.,  1948, Annales d'Astrophysique, \href
  {http://adsabs.harvard.edu/abs/1948AnAp...11..247D} {11, 247}

\makeatother
\end{thebibliography}

% Don't change these lines
\bsp	% typesetting comment
\label{lastpage}
\end{document}